\documentclass[aps,prl,twocolumn,groupedaddress,showpacs,nofootinbib]{revtex4}
\usepackage{graphicx}
\usepackage{times}
\usepackage{amssymb,amsmath}
\def\beq{\begin{equation}}
\def\eeq{\end{equation}}
\def\bey{\begin{eqnarray}}
\def\eey{\end{eqnarray}}
\def\bfig{\begin{figure}}
\def\efig{\end{figure}}

\def\lsim{\mathrel{\raise.3ex\hbox{$<$\kern-.75em\lower1ex\hbox{$\sim$}}}}
\def\gsim{\mathrel{\raise.3ex\hbox{$>$\kern-.75em\lower1ex\hbox{$\sim$}}}}

\def\txt{\text}

\def\order{\mathcal{O}}
\begin{document}

\title{SuperCool Inflation: A Graceful Exit \\from Eternal Inflation at LHC Scales and Below}
\author{Douglas Spolyar}
\affiliation{Center for Particle Astrophysics, Fermi National Accelerator Laboratory, Batavia, IL 60510 USA}
\affiliation{Astronomy and Astrophysics Department and KICP, University of Chicago, Chicago, IL 60637 USA}

\begin{abstract}
In  SuperCool Inflation (SCI), a technically natural and  thermal effect gives a graceful exit to old inflation. 
 The Universe starts off hot and trapped in a false vacuum. The Universe supercools and inflates 
 solving the horizon and flatness problems.
 The inflaton couples to a set of QCD like fermions.  When
 the fermions' non-Abelian gauge group freezes, the Yukawa terms 
generate a tadpole for the inflaton, which removes the barrier.  Inflation ends, and the Universe
 rapidly reheats. The thermal effect is technically natural in the same way that the QCD scale is technically natural.
 In fact, Witten used a similar mechanism  to drive the Electro-Weak (EW) phase transition;
 critically, no scalar field drives inflation,
which allows SCI to avoid eternal inflation and the measure problem.
SCI also works at scales, which can be probed in the lab, and  could be connected to EW symmetry breaking.
Finally, we  introduce a light spectator field 
to generate density perturbations, which match the CMB.
The light field does not affect the inflationary dynamics
and  can potentially generate  non-Gaussianities and 
isocurvature perturbations observable with Planck.

\end{abstract}
\pacs{
98.80.Cq;98.80.Bp;11.30.Qc;12.60.Cn;12.60.Fr
\hspace{0.5cm}FERMILAB-PUB-11-569-A}
\maketitle

\section{Introduction}
\vspace{-10pt}
 In 1981, Guth~\cite{Guth:1980zm} introduced an inflationary phase (old inflation)  to explain the horizon, flatness, and monopole problems.
The Universe begins hot,  cools, and becomes
stuck in a false vacuum and begins to inflate. 
Eventually,
the Universe transitions to the true vacuum and reheats. 
Unfortunately, Guth's model failed due to the Swiss cheese problem or lack of a graceful exit.  
The tunneling rate to the true vacuum  
must be small to generate a sufficient amount 
of inflation, but then inflation never ends.

New inflation~\cite{Linde:1981mu,Albrecht:1982wi}  
sidestepped the Swiss cheese problem by introducing a slowly rolling scalar field to drive inflation.
Slow roll not only solves the standard cosmological problems but can also generate adiabatic density perturbations
consistent with the CMB.
 Regardless, slow roll has two serious generic drawbacks beyond the fine-tuning problem~\cite{Adams:1990pn};
 first, the high scale of inflation (which
prevents  testing inflation in the lab) leads to problems from  trans-Planckian physics  to overclosure from moduli and gravitinos; second (a much  worst
 problem),
eternal inflation
(which leads to the measure problem) 
undermines the predictive power of inflation. Hence, one is interested in alternatives to slow roll
such as cyclic models and ekpyrotic Universes~\cite{Khoury:2001wf}, but these models must contend with a  singular bounce,
which introduces a different can of worms.

   \bfig[t]
\includegraphics[width=0.35\textwidth]{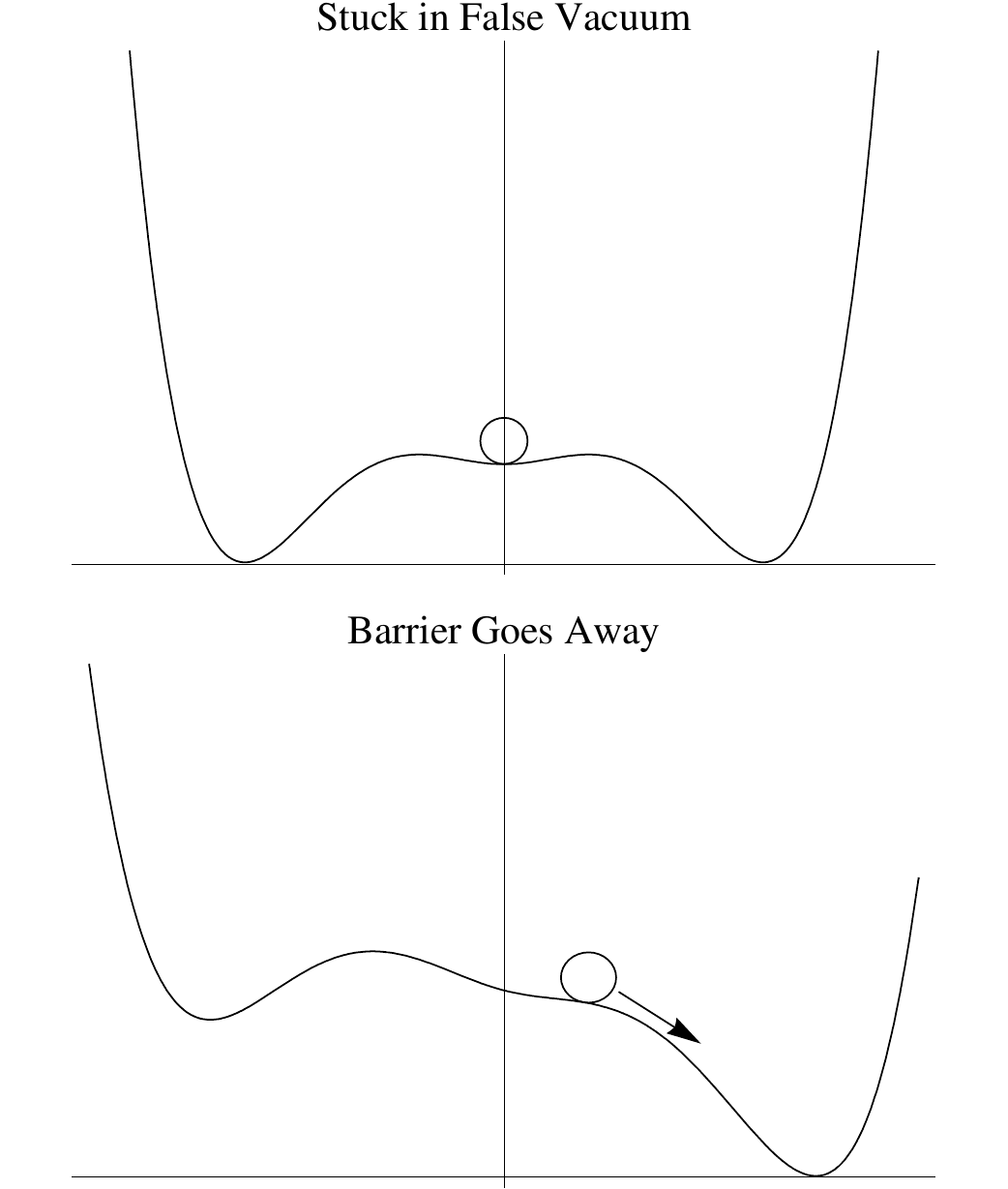}
\centering
\caption{{\bf top--} The Universe is stuck in a false vacuum and supercools.  {\bf bottom--} At T$_c$, a Yukawa term
generates a tadpole term; the barrier goes away.  The Universe then rapidly transitions
to the true vacuum.}
\vspace{-15pt}
\label{apexinf}
\efig

  Instead, we do away with any scalar dynamics to control inflation.  
  A thermal bath (present during old inflation) regulates inflation.
  We introduce a new  
   thermal and technically natural mechanism to generate a graceful exit.
We have dubbed the model SuperCool (SC) inflation since the Universe supercools during inflation and then rapidly transitions
to the true vacuum due to a small perturbation,
in much the same  that a supercooled liquid almost instantaneously freezes if  slightly disturbed. The model in spirit is similar to thermal
inflation~\cite{Lyth:1995ka} except unlike thermal inflation
our model successfully solves the cosmological problems and generates adiabatic density perturbations.
SuperCool Inflation (SCI)  works at the TeV scale and below and avoids eternal inflation.

For simplicity, SuperCool (SC) field is a complex scalar with a Coleman-Weinberg potential
\beq
\label{fp}
\begin{aligned}
&V(\phi)=(1/8)\,\txt{T}^2\,\phi^2+\frac{3g_X^4}{32 \pi^2}|\phi|^4\bigg(\ln\bigg(\frac{|\phi|}{\langle|\phi|\rangle}\bigg)-\frac{1}{4}\bigg)\\
&+\sum y_i \phi\ast (q_R \bar q_L)_i+ \txt{h.c.} +\Lambda^4+\txt{ non-renormalizable}
\end{aligned}
\eeq
 charged under a
U(1)$_X$ gauge group with a charge $g_x$, where $\Lambda^4$ is the standard cosmological tuning, which tunes the cosmological 
constant of the true vacuum to zero. Finite temperature effects generate an effective mass term (T$^2\phi^2$). 
 The SC field  couples to a set of QCD like fermions $q_R \bar q_L$ with a Yukawa coupling $\lambda_i$.
 The SC sector is a simplified version of the standard model with the SC inflaton mapped to the Higgs boson. The
SU(2) weak force accompanied by leptons has been dropped.
The non-renormalizable terms will play an important role in avoiding eternal inflation.
 
 We will discuss Eq.~\ref{fp} in detail in the text but first describe the qualitative behavior of Eq.~\ref{fp}. 
 In the beginning, the Universe is hot and dense, becomes stuck in a false vacuum (Fig.\ \ref{apexinf}{--top}), and 
  inflates solving the horizon and flatness problems. 
  The temperature 
 of the Universe falls exponentially. In section \ref{stabilized}, 
 we show that finite temperature effects stabilize the potential against tunneling during inflation. 
 
 In section \ref{witten}, we introduce a technically natural way to end inflation.
   When the Universe reaches the critical temperature T$_c$,  a non-Abelian gauge group freezes,
 which triggers the end of inflation.
 A set of QCD like fermions charged under the
non-Abelian gauge group  get a vev $\langle \lambda \bar\lambda\rangle\simeq$ T$_c^3$.  The fermions
(which have a  Yukawa coupling to the SC field)
generate a tadpole term for the inflaton potential. The new term removes the 
barrier trapping the scalar field in the false vacuum (Fig.\ \ref{apexinf}{--bottom}).  
We remind the reader that as the temperature of the Universe drops the barrier trapping the false becomes smaller; 
the barrier scales with T.
The field then quickly rolls down to the true vacuum and reheats the Universe.  
The mechanism is technically natural; the end of inflation is determined by the logarithmic
 running of a non-Abelian gauge group. In the same way, the QCD scale is technically natural compared to the Planck scale, 
 which is a difference of $10^{20}$ orders of magnitude and many orders of magnitude more  than
  the temperature difference between when SC inflation begins and ends.

In the case of radiative Electro-Weak (EW) 
symmetry breaking, Weinberg and Guth~\cite{Guth:1981uk} noted that the Universe would become trapped in
the symmetric false vacuum of the Higgs potential to arbitrarily low temperatures.
Witten~\cite{Witten:1980ez} showed that when the QCD vacuum freezes,
 the Yukawa terms generate a tadpole term for the Higgs potential. The new term 
 destabilizes the false vacuum and the Universe transitions to the true vacuum. We take Witten's  
observation and  introduce it as a way to generate a graceful exit for old inflation.

After the field goes to the true vacuum in section \ref{reheating}, 
we show that the Universe rapidly reheats.  We discuss the different ways that the  inflaton can couple
to the standard model.  As an instructive example, we consider kinetic mixing between hypercharge U(1)$_Y$
and  U(1)$_{X}$ which generates a $Z^\prime$.  The SC inflaton then decays into a pair of Z bosons.

 A  high scale  of inflation (order GUT) can be problematic.
 The GUT scale generically results from the 
dual necessity to generate a sufficient number of efolds and  
to generate the correct spectrum of density perturbations~\cite{Adams:1990pn}. 
If the height of the potential is on the order of the GUT scale, then the width frequently needs to be larger
than the Planck scale, at which point we must deal with trans-Planckian physics.  At a practical level,
 non-renormalizable terms suppressed by $m_{pl}$ can become large
 and dangerous~\cite{1997PhRvL..78.1861L}. High scale models also run into overclosure problems 
 from  moduli and gravitinos. 
From a collider perspective, a high scale of inflation is disappointing without a GUT scale collider.
In contrast in section \ref{scale}, we show that SC inflation occurs at the TeV scale and down avoiding the complications of high scale inflation.
  
 Previous authors have proposed inflation at the TeV scale. In fact a model introduced by Turner \& Knox~\cite{Knox:1992iy}
 similarly uses a Coleman Weinberg  (CW) potential (We use a CW potential for the SC field),
  but all of the models  are rolling field models. TeV scale rolling models
  often have difficulties from fine tuning problems~\cite{Lyth:1999ty}
  or require unusual initial conditions~\cite{German:2001tz}.
    SC inflation itself does not suffer from fine-tuning, but our proposed mechanism to 
  generate perturbations (which is similar to a curvaton) suffers from a potential 
   fine-tuning which we discuss in section \ref{perturbations}.
 
  Clearly, we see the interest in connecting TeV scale inflation with EW symmetry breaking 
 or more generally with beyond the Standard Model (SM) building such as hidden valley models~\cite{Strassler:2006im} etc..
  We have not attempted to directly connect SC inflation and EW symmetry breaking, but the path is clear.
  One can write down a gauge invariant dimension 4 renormalizable scalar coupling between the SM Higgs and SC field.
  When the SC field gets a vev, it can generate a negative mass term which induces  spontaneous EW symmetry breaking. 
  We have left detailed model building to future work.  In that spirit, we have written the paper from an effective field theory perspective
  without reference to any particular  fundamental theory which might relate to SUSY, GUTs, etc..

Next, we  turn to eternal inflation in section \ref{eternal}, which is endemic of the slow roll paradigm and troublesome.
  Once one has gone to the trouble of constructing a slow
 roll potential, it has been recently shown that the potential must eternally inflate~\cite{2011PhRvD..84b3511K}.
  In an eternally inflating Universe (a Multiverse for short), 
   most of the Universe is stuck inflating. 
 Regardless, small pockets of the Multiverse will stop inflating and
 could be like our visible Universe or completely different.
 Critically, we need to measure  the relative probability of different pocket Universes to make predictions. 
   After 25 years of extraordinary effort,  no measure predicts a Universe which looks at all like our own.
    For instance, 
 the geometric or light cone measure (introduced by Bousso~\cite{Bousso:2006ev}) predicts 
that time itself will end in the next 5 Billion years~\cite{Bousso:2010yn}.
 We list a few references
 ~\cite{Nomura:2011dt,Harlow:2011az,Garriga:2008ks,
 Bousso:2010im,Linde:2008xf,DeSimone:2008if,Vilenkin:2011yx}
   of  the extensive literature on the measure problem.  

In section \ref{eternal}, we show that SCI avoids  eternal inflation. 
  In general,  any scalar field has two regions in which eternal inflation can crop up: 
  at a hilltop of a potential as with the original new inflation
 model or at large field values as with chaotic inflation.
 The SC inflaton potential has neither of these troublesome points.  
 A non-minimal coupling to gravity or non-renormalizable terms
prevent the large field value case.  At the hilltop points in SCI, 
 the inflaton does not slow roll, in which case there is no
eternal inflation as shown in the Appendix.   In fact, one would need
 to introduce a large tuning to make  the hilltop sufficiently flat to have
   slow roll in the first place.

 We introduce in the last major section \ref{perturbations} a novel way of generating perturbations.
The thermal background which controls SCI  suppresses
 normal density perturbations from a scalar field. We will require a new way to generate perturbations.
 Isocurvature (entropy)  perturbations can generate real adiabatic density perturbations.  Mollerach  showed that if
 a matter component (which has an isocurvature perturbation) decays into radiation, then the matter component's 
 isocurvature perturbation will become a real adiabatic density perturbation~\cite{Mollerach:1989hu}. 
 The curvaton model~\cite{Lyth:2001nq,Lyth:2002my} has implemented Mollerach's original idea.
 We similarly take advantage of Mollerach's mechanism except our model works at much lower inflationary scales compared to the curvaton.
 
 In Section \ref{aulosmechanism}, we introduce  the   aulos
 field (a pseudo Nambu-Goldstone boson). 
Aulos is the Greek root for a flute or reed instrument.
 The field generates real density perturbations, which seed the formation of all known structure in the Universe.
 The ancient Greeks described the motion of celestial objects as the music of the spheres.
Aulos in a similar spirit refers to a source of cosmic harmony.
 After spontaneously breaking the U(1) aulos symmetry in section \ref{aulosmass}, we generate a small technically natural mass
 by explicitly breaking the U(1) symmetry.
 
Then in \ref{aulsoevolution}, we discuss the evolution of the  aulos field.
 During inflation, the  aulos field is Hubble damped, but De Sitter fluctuations
 induce spatial variation of the misalignment angle of the  aulos field. During inflation the mass of the  aulos field ($m_a$) is
 smaller than the Hubble parameter. The  aulos decay constant ($f_a$) is within $\order(10^4)$ of 
 the Hubble parameters.  At the end of inflation, both $m_a$ and $f_a$ grow and become substantially larger than H,
 which transfers energy from the inflaton field into the aulos field. 
 The  aulos field begins to
 oscillate, which generates a cold condensate of the aulions. Now, the spatial variation of the misalignment 
 angle of the  aulos field induces an isocurvature perturbation. The  aulos field then
  decays into radiation and generates real density perturbations.
 
  Next, we show in section \ref{signatures} that
   the  aulos mechanism can produce the perturbations seen in the CMB
    and  potentially generate some novel features.  
  The  aulos field is very similar to the curvaton scenario except the aulos mechanism works with inflationary scales much smaller than $10^9$ GeV.  
  Hence, many of the curvaton features
   apply to the  aulos scenario. First, the  aulos mechanism 
  generates the scale of perturbations seen in the CMB and the spectral tilt 
  of the power spectrum. Second,   the very low scale of inflation
  suppresses  primordial gravity B--modes. 
    The aulos field can also potentially generate levels of  non-Gaussianities
    and isocurvature perturbations, which are observable with Planck and other future missions. 
   In section \ref{collider}, we then connect the cosmological parameters to  the
   parameters of the aulos field which could be measured in a lab.
   
We would like to point out  that the aulos mechanism potentially has  applications beyond providing perturbations
during inflation. We define  an aulos field as any field
    which has  a small decay constant ( spontaneous symmetry  breaking scale ) and/or mass
   during inflation. At the end of inflation, the field has  a large decay constant and/or mass.
We will discuss some potential applications in the conclusions from variations in dark energy to spatial variation in $\alpha$~\cite{doug1}.

It is important to emphasize despite the many different issues discussed in the paper that the underlying idea is simple. If we can
free ourselves of requiring the inflaton to generate perturbations, we can do away with slow roll altogether and the
many complications which result from slow roll from technical concerns about fine-tuning to more prosaic concerns about the measure problem.
We can also think seriously about low scale inflation.

 \section{SuperCool Inflation}\label{SCinflation}

\bfig[t]
\includegraphics[width=0.35\textwidth]{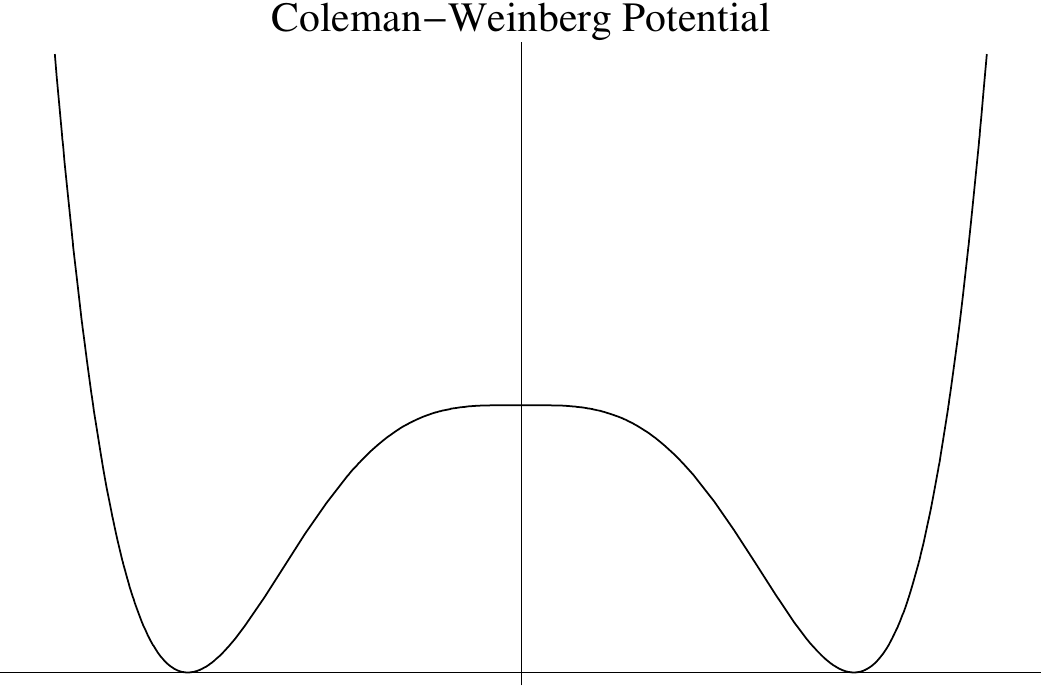}
\centering
\caption{Coleman Weinberg potential for the SC field $\phi$  (See Eq.~\ref{CW}), at zero temperature. }
\vspace{-15pt}
\label{CWfig}
\efig

As our starting point, the SuperCool (SC) field $\phi$ has a Coleman Weinberg (CW) potential
 (See Fig.~\ref{CW}) at zero temperature.  Furthermore, SC field is a complex scalar, which is charged under a 
U(1)$_X$ Abelian group.  We have enforced classical scale invariance on the potential ( known as the ``no bare mass" 
condition)\footnote{The naturalness of Eq.~\ref{CWcondition} is open to question, but Gildener and Weinberg have argued that 
the ``no bare mass" condition is  natural~\cite{Gildener:1976ih}.  In addition,  CW potential have been  used
extensively for EW symmetry breaking and slow roll inflation. 
At the minimum, we take classical scale invariance (``no bare mass" condition) as an interesting hypothesis 
and continue.} such that 
\beq
\label{CWcondition}
\frac{d^2\txt{V}(\phi)}{d\phi^2}\bigg\vert_{\phi=0}=0.
\eeq

Classically, the SC field is simply of the form $\lambda\phi^4$ and is  scale invariant. Quantum corrections modify the classical potential 
breaking the scale invariance. Then, dimensional transmutation generates a nonzero vacuum expectation value  $\langle\phi\rangle$ i.e.\ vev.
At one loop, the potential~\cite{Coleman:1973jx} is
\beq
\label{CW}
\txt{V}(\phi)=\frac{3g_X^4}{32 \pi^2}|\phi|^4\bigg(\ln\bigg(\frac{|\phi|}{\langle|\phi|\rangle}\bigg)-\frac{1}{4}\bigg)+ \Lambda^4
\eeq
 where $g_X=0.4$ is the charge of the scalar field.   $\Lambda$ is the usual cosmological tuning, which will generate the 
vacuum energy during inflation V$(0)=\Lambda^4$ and sets the true vacuum to zero V$(\langle \phi\rangle)=0$.
We will take $\Lambda\simeq1$ TeV and $\langle|\phi|\rangle\simeq10$ TeV to be concrete.
Clearly, the vev $\langle\phi\rangle$ and the scale of inflation are connected for a given $g_X$.
The scale of inflation i.e.\ $\Lambda$ can be larger or smaller than 1 TeV;
 we will comment later. So far, we have neglected temperature effects.

\begin{figure*}
\includegraphics[width=.7\textwidth]{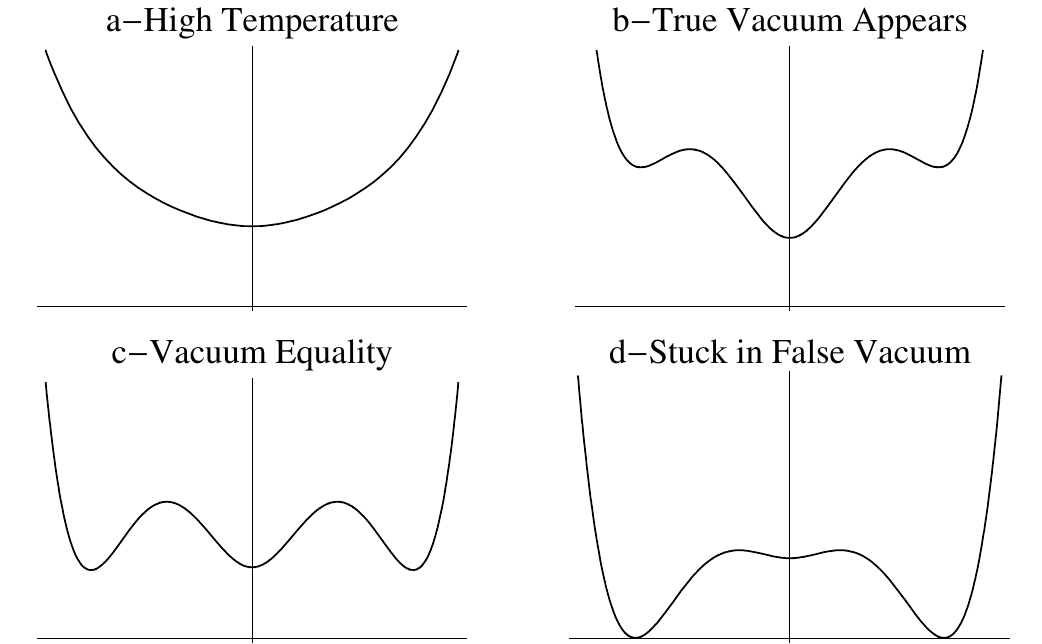}
\centering
\caption{The effective potential at different temperatures:
 {\bf a}--At high temperatures, only one vacuum state exist. 
{\bf b}--Eventually, other vacuum states appear.
{\bf c}--At the transition temperature, the Universe will be in the symmetric minima.
 {\bf d}--As the temperature drops, the symmetric vacuum remains metastable.}
\label{tempfig}
\end{figure*}

\subsection{Stabilized Potential}\label{stabilized} 

At a nonzero temperature, finite temperature effects stabilize the false vacuum of the SC potential.  
Near the origin of $\phi$, we can approximate
the finite temperature part of the effective potential with
\beq
V(\txt{T}, \phi)=\order(1)\txt{T}^4+(1/8)\,\txt{T}^2\,\phi^2+\order(\phi^4)
\eeq
 (\cite{Witten:1980ez,Dolan:1973qd,Weinberg:1974hy}).  At very high temperatures (Fig.\ref{tempfig}-a), the symmetric minimum is the only 
vacuum state and the Universe sits at $\phi=0$. As 
the temperature of the Universe drops (Fig.\ref{tempfig}-b),  the true vacuum appears near $\phi=\langle\phi\rangle$.
The potential evolves only gradually from high temperatures. As the Universe continues 
to cool reaching the transition temperature (Fig.\ref{tempfig}-c), 
the Universe will still be in the symmetric minima~\cite{Guth:1980zk,Witten:1980ez}. 
 At temperatures below the transition temperature (Fig.\ref{tempfig}-d), thermal tunneling can allow the field
to transition to the true vacuum, but the rate is exponentially small. 
We will follow an argument first given by Witten~\cite{Witten:1980ez}.

The tunneling rate $\Gamma$ will be dominated by the O(3) thermal instanton S$_3$~\cite{Linde:1981zj}.
We have also considered an O(4) instanton~\cite{Linde:1981zj}
and a Hawking-Moss instanton~\cite{Hawking:1981fz}, which are subdominant.
\beq
\label{tunnelingrate}
\Gamma\sim \txt{T}^4 \bigg(\frac{\txt{S}_3(\txt{T},\phi)}{2\pi\, \txt{T}}\bigg)^{3/2} \exp[{-\frac{\txt{S}_3}{\txt{T}}}]
\eeq
where T is the temperature of the Universe. Near the origin and  the barrier (the part of the potential relevant
for tunneling), we can transform Eq.~\ref{CW} with 
a trick invented by Witten~\cite{Witten:1980ez}
\footnote{Witten shows that near the origin and the barrier, which is the relevant  part of 
the potential for tunneling that
$\ln(\phi/\langle\phi\rangle)-1/4 \longrightarrow-\ln(\txt{M}/\txt{T})+\mathcal{O}(1)$.}
\beq
\label{approx}
\txt{V}(\txt{T},\phi)=\frac{\txt{T}^2}{8}|\phi|^2-\frac{3 g_X^4}{32 \pi^4}|\phi|^4\bigg(\ln\bigg(\frac{\txt{M}_{X}}{\txt{T}}\bigg)
+\mathcal{O}(1)\bigg)
\eeq
where M$_{X}=g_X\langle\phi\rangle$ is the mass of the U(1)$_X$ gauge boson.  With Eq.~\ref{CW} now in the
form of Eq.~\ref{approx},
we have an exact solution of the O(3) instanton. 
\beq
\label{S_3}
\txt{S}_3=\frac{4 \pi^2}{3} \frac{\txt{T}}{g_X^3\ln(\txt{M}_X/\txt{T})} 19
\eeq
where the factor of 19 is a geometric factor. 

In SCI, the Universe is stable against tunneling to the true vacuum. At most, only some
small part of the Universe can transition to the true vacuum. 
Guth and Weinberg~\cite{Guth:1982pn} have shown that the tunneling rate (per unit time per unit volume) $\Gamma$ 
compared to the  Hubble 4-Volume  
\beq
\label{trate}
\beta=\Gamma/\txt{H}^4\gsim9/4\pi=\beta_c, 
\eeq
must be larger than $\beta_c$ for the Universe to transition from the false to the true vacuum. 
We show that $\beta$ for the SC field is much smaller than $\beta_c$ during inflation. Hence, the Universe is stuck in the false vacuum state.
 Rapid tunneling  to the true 
vacuum  can only occur when S$_3\rightarrow0$.     Upon inspection,  only as $g_X$ becomes large or
 T goes to zero does S$_3\rightarrow0$, but we show that in either case tunneling will still be extremely small.  Hence, a new mechanism will
 be needed to generate a graceful exit.
 
 First, $g_X$ never becomes large and stays  perturbatively  small as the temperature drops
from the TeV scale down to a fraction of an electron volt (T$\leq 10^{-2}$eV).  At which point in the next section, we show 
that the Witten mechanism can generate a graceful exit.  More generally if the SC field was charged under a non-Abelian gauge group, 
we would need to be more careful (See \cite{PhysRevD.24.1699}).

Second as the temperature drops, the logarithm $\ln(\txt{T}/\txt{M}_{X})$ in Eq.~\ref{S_3} blows up. Regardless $\beta$ (for the inflaton potential)
 is much less than 
$\beta_c$ during inflation, since
the pre-factor in Eq.~\ref{tunnelingrate}
 (which scales like T$^4$) goes to zero sufficiently fast to counteract the Log factor in the exponential. 
 Furthermore, $\beta$ is sufficiently small to avoid constraints from BBN and CMB~\cite{Turner:1992tz}.

   In addition, a technical concern could lead to a much larger tunneling rate.
 At the origin of the SC potential,  
     the one loop perturbative calculation given in Eq.~\ref{CW} becomes no perturbative. In principle, the actual potential could be very different leading
     to a much larger tunneling rate.  The worry is unfounded.  
Subsequently, a non-perturbative calculation~\cite{1997MPLA...12.2287L} has 
been preformed which verifies that Eq.~\ref{CW} is still correct.  Hence, the
 Universe is safe from transitioning from the false vacuum to the true vacuum during inflation. In fact, the Universe never tunnels for even
  arbitrarily low temperatures.  For inflation to end, we will need to introduce new physics.

\subsection{ Ending Inflation (Witten Mechanism):}\label{witten}
Witten's mechanism generates a graceful exit and ends inflation.
We implement the Witten mechanism by introducing a set of QCD like fermions charged under a
non-Abelian gauge group with a Yukawa coupling to the SC field V$_{\txt{y}}(\phi)=\sum y_i \phi\ast (q_R \bar q_L)_i+$ h.c. 
where $y_i$ is the Yukawa coupling for 
respectively the right and left-handed fermions $q_R$ and $q_L$. 
The SC field $\phi$ is a singlet under the non-Abelian gauge group.   Only some of the right handed quarks and left handed
 quarks are charged under the  U(1)$_X$ such that  the Yukawa terms are gauge invariant.  We also must 
 be careful how we assign U(1)$_X$ charges  to the quark like fermions to ensure that the theory is anomaly free. 
  As with QCD, we can pick
a sufficiently small coupling constant such that the theory will only become strongly coupled at a low scale.
Finally, the fermion loop corrections have not been included in Eq.~\ref{CW}. Qualitatively the evolution of the aulos field
remains unchanged unless there are a large number of fermions with large Yukawa couplings.\footnote{
Witten used the same approximation and  found a similar conclusion~\cite{Witten:1980ez}.}

When the temperature of the Universe drops below the strong coupling 
scale $\Lambda_S$ of the non-Abelian gauge group, the vacuum
of the non-Abelian gauge group freezes, and the fermions gain a vev 
\beq
\label{break}
\txt{V}_y(\phi)\rightarrow\bigg(\sum y_i \langle q_R \bar q_L\rangle_i\bigg)\ast \phi+\txt{h.c.}=\epsilon \ast\phi+\txt{h.c.}.
\eeq
 The vacuum seizes, which dynamically breaks  any global symmetries
possessed by the quark like fermions $q_R q_L$, and the U(1)$_X$ gauge 
symmetry.\footnote{ 
As Witten~\cite{Witten:1980ez} pointed out, a more rigorous analysis  would replace 
$\langle q_R\bar q_L\rangle(x)=$Q$(x)$ with an order parameter  to describe symmetry breaking
in order to more carefully account for gauge invariance, but the underlying analysis would not change.
}  
If some of the $y_i\simeq1$, then
 $\epsilon\simeq\Lambda_S^3\simeq \txt{T}_c^3$, where T$_c$ is the temperature at which the
  vacuum of non-Abelian gauge group freezes. As we show below, the barrier trapping the SC field in the false vacuum goes away
  once the non-Abelian gauge group freezes.

\vspace{10pt}
{\bf Concrete Example}
\vspace{10pt}

We now work through a concrete case of SCI with $\Lambda\simeq1$TeV and
 $\langle|\phi|\rangle\simeq10$ TeV.
Initially, the Universe is radiation dominated.  The Universe then cools and becomes stuck in the false vacuum. 
The energy density of Universe becomes dominated by vacuum energy.
The Universe begins superluminal expansion once the temperature falls
below a few hundred GeV (N.B.\ The precise temperature when superluminal growth begins depends upon the 
number of relativistic degrees of freedom.  At temperatures between 100GeV to 1 TeV, there 
are $\order(100)$ relativistic degrees  of freedom
from the standard model.  In which case, superluminal expansion begins once the temperature falls to $\sim300$ GeV).   
As a conservative estimate, we will assume that inflation really only starts once the temperature of the Universe is a 100 GeV.
 Inflation must last for at least 30 efolds to solve the horizon and flatness problems (See Eq.~\ref{efolds}). 
 After 30 efolds, the temperature of the
  Universe drops to $\sim10^{-2}$~eV and the non-Abelian gauge group freezes. See Eq.~\ref{break}.

Once the non-Abelian gauge group freezes, the barrier goes away.  Near the origin,
Eq.\ref{approx} together with Eq.\ref{break}   has the form 
\beq
 (\epsilon\phi+ \txt{h.c.})+\frac{1}{2}\,\tilde m^2|\phi|^2-\frac{1}{4}\,\tilde\lambda |\phi|^4
\eeq
 (See Fig.\ \ref{apexinf}).
We note that the linear term will destabilize the meta-stable vacuum state ($\phi\simeq0$ )  by eliminating
the barrier if
\beq
\label{dest}
|\epsilon|\geq\frac{2}{3^{3/2}}\frac{\tilde m^3}{\tilde\lambda^{1/2}}.
\eeq
Upon substituting in values for $\epsilon$, $\tilde m$, and $\tilde\lambda$, we see that as the temperature drops so does m; while, $\tilde\lambda$
increases. In Fig.\ref{apexinf}, the barrier trapping the inflaton at the origin goes away once T $\simeq$ T$_c\sim10^{-2}$ eV, 
with   $g_X=0.4$,
M$_{\chi}\simeq$ 4 TeV and 
T$_c\simeq\Lambda_S\simeq10^{-2}$ eV. In sum, the Universe goes through 30 efolds of inflation solving the 
flatness and horizon problems, the barrier  goes away, and  
the  SC field goes 
rapidly to the true vacuum. The Universe then reheats as discussed below. 

\subsection{Reheating}\label{reheating}

Once  the barrier goes away, the field quickly goes to the 
true minimum, where it may oscillate and  decay into standard model particles.  Reheating can be virtually
instantaneous for the decay rate $\Gamma_\phi \gg$H. 
We might have a complicated decay process into standard model particles. For instance,
the SC field might decay into particles charged under U(1)$_X$ which subsequently decay into standard 
model particles.
If the inflaton only partially decays into standard model particles, 
then the SC sector might be related to dark matter.  We will instead treat the case 
where $\phi$ decays  only into standard model particles and in particular Z bosons.  
The choice is idiosyncratic; many other reasonable 
decay routes exist (for instance, the SC field could decay into Higgs bosons).

 In general, there are only 3 couplings between a 
hidden sector and the standard model which are renormalizable and gauge invariant: a vector coupling
to hypercharge $B_{\mu \nu}$ (kinetic mixing), a scalar coupling to the Higgs HH$^\dagger$ , 
and a spinor coupling with a Higgs-neutrino operator HL~\cite{2009arXiv0909.4541W}.  We will focus on the kinetic 
mixing~\cite{Holdom:1985ag} between hypercharge $B_{\mu\nu}$ and the U(1)$_X$ SC gauge boson $X_{\mu\nu}$.
We have, thus, expanded the standard model $\rightarrow$ SU(3)$\times$SU(2)$_{\txt{L}}\times$U(1)$_Y\times$U(1)$_X$.
In which case,  the standard model will contribute to the CW potential Eq.~\ref{CW}. If the EW phase transition occurs during
inflation, then there will be a threshold correction to the CW potential.
  The renormalization condition Eq.~\ref{CWcondition} can be 
maintained by matching the renormalization group flow equations as it runs from the IR  up to the threshold 
and from the UV down to the threshold.  The IR and UV are then necessarily sensitive to each other,
which only highlights the mysterious nature of the  ``no bare mass" condition.  

In the case of kinetic mixing between hypercharge U(1)$_Y$ $B^{\mu\nu}$ 
and the X boson U(1)$_X$ $X^{\mu\nu}$, the kinetic energy terms go like
\beq
\label{ke}
\mathcal{L}_{\txt{KE}}^{BX}=-\frac{1}{4}B_{\mu\nu}B^{\mu\nu}-\frac{1}{4}X_{\mu\nu}X^{\mu\nu}+\frac{\chi}{2}B_{\mu\nu} X^{\mu\nu}
\eeq
where $\chi$ (at an effective level) is an arbitrary mixing parameter 
and can take any value (which is how we treat $\chi$ for the remainder of the paper).
 N.B.\  in some top down approaches  the parameter $\chi$ can arise from integrating out vector like fermions
charged under both the hidden U(1)$_X$ and hypercharge~\cite{Holdom:1985ag}, which 
gives $\chi\sim10^{-2}$.

Upon diagonalizing the kinetic term Eq.~\ref{ke} with a GL(2,R) rotation and then diagonalizing the gauge
boson mass matrix with an O(3) rotation, we can write $X_\mu$ in terms of the mass eigenstates of the standard 
model Z boson and a Z$^\prime$~\cite{2009arXiv0909.4541W}. We now have the coupling of the SC field $\phi$ to the Z boson.

We can, then, determine the decay rate of the SC field into Z bosons.
When the mass of the  
U(1)$_X$ gauge boson M$_{X}$ is large
compared to the unmixed Z boson mass M$_{Z_0}$ (the $0$ refers to the field before mixing)
 or $\chi$ is small, the decay rate of the SC
inflaton into a pair of Z bosons~(\cite{Lee:1977yc}, ~\cite{Lee:1977eg}, and~\cite{Gunion:1989we}) is then
\beq
\label{Apexdecay}
\Gamma_{\phi\rightarrow \txt{ZZ}}=\frac{\pi}{6} M_\phi (g_X\, \sin\theta_w \eta \Delta z)^4
\frac{\sqrt{1-x}}{x^2}(3x^2-4x+4)
\eeq
where M$_\phi$ ($\sim300$ GeV) is the mass of the SC inflaton, $\sin\theta_w$ is the Weinberg angle, $\Delta z=(M_{Z_0}/M_{X})^2$,
 $x=4M_Z^2/M_\phi^2$, $\eta=\chi/\sqrt{1-\chi^2}$ and $g_X=0.4$. We find that 
 $\Gamma_{\phi\rightarrow \txt{ZZ}}= 0.03$ eV$\gg$ H $\simeq10^{-4}$ eV,
  where we have taken $\chi=.9$ and $M_{Z^\prime}\simeq M_X$ ($\sim4$ TeV). The field  will predominantly decay into Z bosons,
  if  the masses of all other particles which couple to $\phi$ are more massive than $M_\phi$.  
  For instance the QCD like fermions have Yukawa couplings which are
$\order$(1) and have a mass $\sim$10 TeV.
  
 We have assumed a large mixing $\chi$ between the $Z_0$ and $X$ boson but the large difference in the masses suppresses the decay rate.
 One could imagine a slightly less massive X particle which would lead to a much larger decay rate. Regardless, the Universe rapidly reheats
 converting the vacuum energy into radiation.  With 100 relativistic degrees of freedom, 
 the Universe then reheats to a  temperature $\simeq500$ GeV,
 which is sufficiently high to do EW baryogenesis~\cite{Wainwright:2009mq}.
  If we had a smaller mixing parameter,  we could then have a smaller decay constant and
 a lower reheat temperature.  
 
 The SC field can avoid collider constraints. 
 The Z$^\prime$ is heavy enough to avoid present collider bounds~\cite{2011PhRvD..83g5012A,Appelquist:2002mw}.  The coupling
 of $\phi$ to the standard model is strongly suppressed in this case due to large mass ratio of Z to Z$^\prime$.
 Hence, the toy model is consistent with collider bounds. In a future paper, we will consider collider searches for the SC inflaton~\cite{doug1}.

 \subsection{Scale of Inflation}\label{scale}
 
 Cosmological considerations can place an upper and lower limit on the scale of SCI.
 On the low end, any inflation model must satisfy Big Bang Nucleosynthesis (BBN) constraints. 
 In which case, the Universe then reheats to at least a few MeV. 
 Baryogenesis could potentially, also, place constraints on the minimal scale of inflation. 
   If the scale of inflation and reheat temperature
   is above 100 GeV, then one can use EW baryogenesis or leptogenesis to generate a baryon 
   asymmetry~\cite{Wainwright:2009mq}.  At present, we know of no published models of  baryogenesis which work
   with the scale of inflation below 100 GeV.
   Thus, our present lack of imagination with regards to baryogenesis will require that inflation occurs above 100 GeV.

 An aulos field might get around the above argument.  As defined in the introduction, an aulos field is any field which has a small mass
  during inflation and a large mass at the end of inflation.  Later in the paper, we will use the aulos mechanism to generate density perturbations.
  In a different direction, the aulos mechanism in conjunction with the Affleck-Dine Mechanism (ADM)~\cite{Affleck:1984fy}
  could also be used to generate a baryon asymmetry  even if the scale of inflation is below 100 GeV.  We would like to emphasize
  that the field generating a baryon asymmetry and the field generating density perturbations are not necessarily one and the same.
  
The Affleck-Dine Mechanism (ADM) can occur with a reheat temperature as low as a few MeV, but 
  ADM without the   aulos mechanism requires that the scale of inflation be above $10^9$ GeV~\cite{Dine:2003ax,Banks:1996ea}.
  The constraint on the scale of inflation from ADM
  follows from 2 requirements: 1-- the mass of the field carrying baryon number must be larger than the mass of the proton
 when the field decays.  2-- the field must also be Hubble-damped 
  during inflation.  
  
  An  aulos field carrying baryon number can satisfy the 2 ADM requirements and could be used to do baryogenesis at a scale of inflation below 
  $10^9$ GeV.
  The mass of the  aulos field is small during inflation so the field is Hubble damped satisfying
  the first requirement regardless of the inflationary scale.   At the end of inflation, the mass of the field becomes large satisfying
  the second requirement. The field begins to oscillate and generate baryon number via ADM.
  In principle, one can have baryogenesis for a very low inflationary scale (potentially  MeV scale). 
    We will leave detailed model building for a future paper~\cite{doug1}.
  Regardless from observational constraints, the scale of inflation must still be larger than a few MeV.
  
 On the other end of the scale, SCI most naturally occurs for $\Lambda\lesssim100$ 
 TeV due to the appearance of non-renormalizable terms and 
  the number of efolds necessary to solve the horizon and flatness problems.
For instance, we have so far neglected a term which arises in general relativity 
  \beq
  \label{gcoupling}
  -\frac{1}{2} \xi R|\phi|^2,
  \eeq
   where $R=-32\pi G\Lambda^4$. $G$ is Newton's constant and $\Lambda^4$ is the vacuum energy during inflation. $\xi$
  characterizes the coupling of the SC field $\phi$ to gravity. In the minimally coupled case ($\xi=0$),
  there is no constraint on the scale of inflation. The scale can be arbitrarily large. 
  As  another example, if the ``no bare mass" hypothesis is true, then it
  is sensible to believe that the SC field is classically conformally invariant
  which implies that $\xi=1/6$~\cite{Callan:1970ze}~\cite{Abbott:1981rg}.
  We note that if we include loop corrections $\xi$ is a renormalizable coupling constant that runs to a fix point $\xi=1/6$ in the IR but 
 can have a negative value in the UV~\cite{Hill199221}.

  Regardless, Eq.~\ref{gcoupling} can be problematic.
  If $\xi>0$,  Eq.~\ref{gcoupling}  stabilizes the false vacuum,  in which case inflation could never end.
  If $\xi<0$, Eq.~\ref{gcoupling} could destabilize the false vacuum state, in which case inflation could end too soon.
 Hence, we require that inflation ends before Eq.~\ref{gcoupling} becomes important 
  \beq
  \label{gcrit}
  \txt{T}_c^2\gtrsim |\xi R|\propto |\xi| G \Lambda^4
  \eeq
  where T$_c$ is the critical temperature  when inflation ends (See Eq.~\ref{dest}).
  
  The number of efolds of inflation necessary to account for the horizon problem goes like
  \cite{1990eaun.book.....K} 
  \beq
 \label{efolds}
\txt{N}_{min}=30 + \frac{1}{3} \left( 2 \log \left[ \frac{\Lambda}{1\text{ TeV}}\right] 
+ \log\left[\frac{\txt{T}_{RH}}{1\text{ TeV}}\right] 
 - \log\left[\gamma \right]\right)
\eeq
 where $\Lambda$ is the scale of inflation, T$_{RH}$ is the reheat temperature at the end of inflation, 
 and $\gamma$ is the amount of 
 any subsequent entropy production after inflation and reheating. 
Also, the number of efolds necessary to resolve the flatness problem  is  similar to the number
 of efolds necessary to solve the horizon problem (See \cite{1990eaun.book.....K} Chapter on inflation). 
 We will require that the number of efolds of inflation be greater than or equal to  Eq.~\ref{efolds}.
 
 Tension between Eq.~\ref{gcrit} and Eq.~\ref{efolds} places an upper limit on the scale of SCI.
   The necessary number of efolds increases with the scale of 
 inflation (first term of Eq.~\ref{efolds}),  but Eq.~\ref{gcrit}  pushes up T$_c$ with an increasing scale 
 of inflation.
 We can only satisfy both Eq.~\ref{gcrit} and Eq.~\ref{efolds} if $\Lambda\lesssim100$ TeV for $\xi\sim\order(1)$.
 
 SCI comfortably occurs  between 100 TeV  and 100 GeV (also potentially even the MeV scale),
  but inflation at the TeV scale would certainly be
 interesting.   In many models beyond the standard model,
 new machinery at the TeV scale is introduced to explain EW  symmetry breaking: 
   gauge mediation, little Higgs models, large extra dimensions etc.. 
 Theoretically, it is compelling to connect inflation to the dynamics which drive EW symmetry breaking.  
 Second,  LHC and future colliders can effectively probe the TeV scale.  Inflation could become a laboratory science.
   At the moment, we can only probe inflation indirectly through cosmology.
Finally,  the low scale of inflation inherently avoids many pitfalls of high scale inflation such as  overclosure from gravitinos to moduli
and trans-Planckian physics.
 
 \section{No Eternal Inflation}\label{eternal}
    
Virtually, all known models of inflation suffer from eternal inflation such as chaotic inflation~\cite{1986PhLB..175..395L} and new inflation
  (See Fig.\ \ref{eternfig}).
  There are a few exceptions such as~\cite{Baumann:2007np} which require a large amount of fine-tuning.  
 More generally as noted by Guth, most rolling models without eternal inflation
  are pretty contrived from a field theoretical perspective~\cite{Guth:2000ka}. 
 
 As a simple example of eternal inflation consider chaotic inflation, which occurs at large field values.   
 A scalar field is displaced 
from the true minimum. The potential of the scalar field V$(\phi)$ is sufficiently flat such that the field rolls very 
slowly to the true vacuum state. 
 As the Universe rolls down, quantum fluctuations of the field
 can cause a small patch of the Universe to go up the potential, which cause the fluctuation to grow (See Fig.\ \ref{eternfig}). A small patch can
continue to fluctuate up the potential until the quantum fluctuations of the field become order one i.e. $\Delta\rho/\rho \sim 1$,
at which point the patch can no longer roll back down.  The patch begins to exponentially expand.
 At which point, the Universe begins to inflate eternally and only an exponentially small fraction of the 
 Universe is not eternally inflating (See~\cite{Guth:2000ka} for an introduction to eternal inflation). These small pocket Universes
 could evolve into a Universe which looks like ours or could be very different.

   If we had  a full picture of the eternally inflating Universe or Multiverse for short,
   we could generate a probability distribution for the properties of different pocket Universes, such as
 the cosmological constant, the amount of dark matter in the 
 Universe etc..   Then in fact, eternal inflation could be able to make various predictions and be tested.
 
 Sadly, counting in an infinite Universe proves difficult.
  To quote Alan Guth, ``In an eternally inflating Universe, 
  anything that can happen will happen; in fact, it will happen an infinite number of times. 
  Thus, the question of what is possible becomes trivial -- anything is possible, 
  unless it violates some absolute conservation law. 
  To extract predictions from the theory, we must therefore learn to 
  distinguish the probable from the improbable."~\cite{Guth:2000ka}.
Hence at worst, we have given up any predictive power of inflation 
since anything and everything is possible, but if one had
 a good way to count the relative occurrence of different pocket Universes, 
we could regain the predictive power of inflation.

At present, there is no agreed method of determining the relative probability of different pocket Universes,
or in other words  we have no sensible measure of the Multiverse.
In fact, eternal inflation and various measures so devised  have instead led to a series of problems
such as the youngness problem, Boltzmann brains, etc., or make predictions which are wholly counterintuitive.
 For instance, 
the geometric or light cone  measure (introduced by Bousso~\cite{Bousso:2006ev}) predicts 
that time itself will end in the next 5 Billion years~\cite{Bousso:2010yn}.
(See \cite{Freivogel:2011eg} for a recent review). One might be interested  in coming up with an inflationary model which avoids
 eternal inflation in the first place.

  \bfig
 \includegraphics[width=0.5\textwidth]{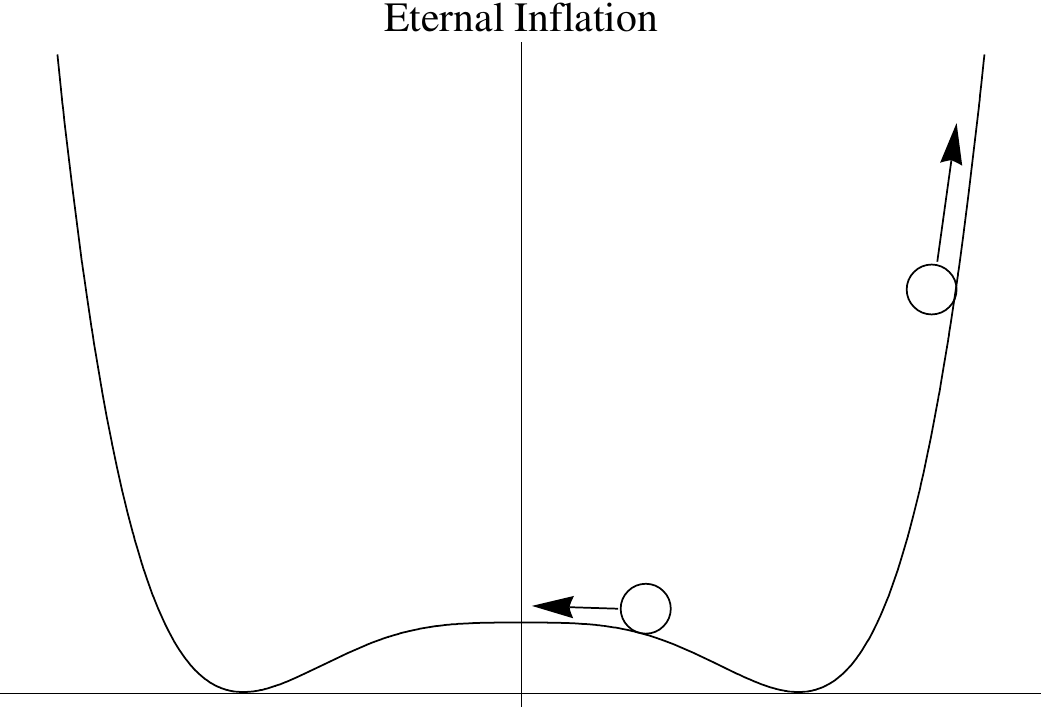}
 \centering
 \caption{Eternal inflation occurs at the origin of slowly rolling fields (as in new inflation)
  and at large field values (as with chaotic inflation).
 Quantum fluctuations prevent the field from rolling to the true vacuum as represented with the arrows.}
 \label{eternfig}
 \efig

 Eternal inflation occurs when 
\beq
\label{EIcond}
\frac{\delta\rho}{\rho}\gsim1 \rightarrow \frac{V(\phi)^{3/2}}{m_{pl}^3}\gsim |V^{\prime}(\phi)|
\eeq
where $m_{pl}=2.4\times10^{18}$ GeV is the reduced Planck mass, $V(\phi)$ is the SC field but more generally is 
any scalar potential. The prime refers to partial differentiation with respect to $\phi$.
 For any scalar field, there are potentially two problematic field values.
 First in the large field case  ($\phi\rightarrow\infty$), the field could undergo chaotic eternal inflation as discussed above.  Second near 
 the hilltop of the potential (a local maximum-- $V^\prime(\phi)=0$), a scalar field can also undergo eternal inflation as with the original new inflation models.  

   Regardless, SCI actually avoids eternal inflation. We consider both the hilltop and the chaotic cases.
  In the hilltop case  ($V^\prime(\phi)=0$), we show in the Appendix, that  if
  \beq
  \label{NoEI}
  \sqrt{|V^{\prime\prime}(\phi)|}>3\txt{H},
  \eeq
 then there is no eternal inflation. Conversely if 3H$>\sqrt{|V^{\prime\prime}(\phi)|}$, then
  there will be eternal inflation.
 There is a hilltop point near the origin of the SC field ($\phi\simeq0$). We note that the tadpole term shifts the hilltop point away from the origin.
   By plugging in numbers, it is clear
 that there is no eternal inflation near the origin since slow roll is violated.
 
 SCI can avoid chaotic or large field value eternal inflation.
 In the large field case for the SC potential
   Eq.~\ref{CW},  we violate Eq.~\ref{EIcond} once  $\phi >m_{pl}$.   We have so far not discussed the non-renormalizable
 terms given in Eq.\ref{fp}.   The SC potential $V(\phi)$ reaches a maximum before $\phi >m_{pl}$ by including a non-renormalizable term.
 
 Non-renormalizable terms are a necessary component of an effective field theory.
 For instance, a field which is non-minimally coupled to gravity has just such a term.
  Non-minimal coupling was discussed in the previous section.
 More generally, non-renormalizable terms can arise, when embedding a effective field theory in a more fundamental
 theory. Non-renormalizable terms will go like $(\pm)\lambda_n \Lambda_c^{-n} \phi^{4+n}$, where $\Lambda_c$ 
 is the ultra-violet cut off and $\lambda_n$ is positive.   
  We now include a non-renormalizable term in the potential of the SC field
 Eq.~\ref{CW}
 \beq
 \label{CWredux}
V(\phi)=
 \frac{\lambda}{4}|\phi|^4\bigg(\ln\bigg(\frac{|\phi|}{\langle\phi\rangle}\bigg)-\frac{1}{4}\bigg)
 -\frac{\lambda_{2n}}{2n+4}\frac{|\phi|^{2n+4}}{\Lambda_c^{2n}}
\eeq 
where $\lambda= 3g_X^4/8\pi^2$ (See Fig.\ \ref{noeternfig}).
For simplicity, we have neglected the cosmological tuning parameter in Eq.~\ref{CW}
 (N.B.\ neglecting $\Lambda\ll\Lambda_c$ will not alter   our results ) and the tadpole term which is irrelevant at large field values.
By inspection of Eq.~\ref{CWredux}, there exists a hilltop point $\phi_0$ where $V(\phi_0)$ is a maximum of the potential. 

 \bfig
 \includegraphics[width=0.5\textwidth]{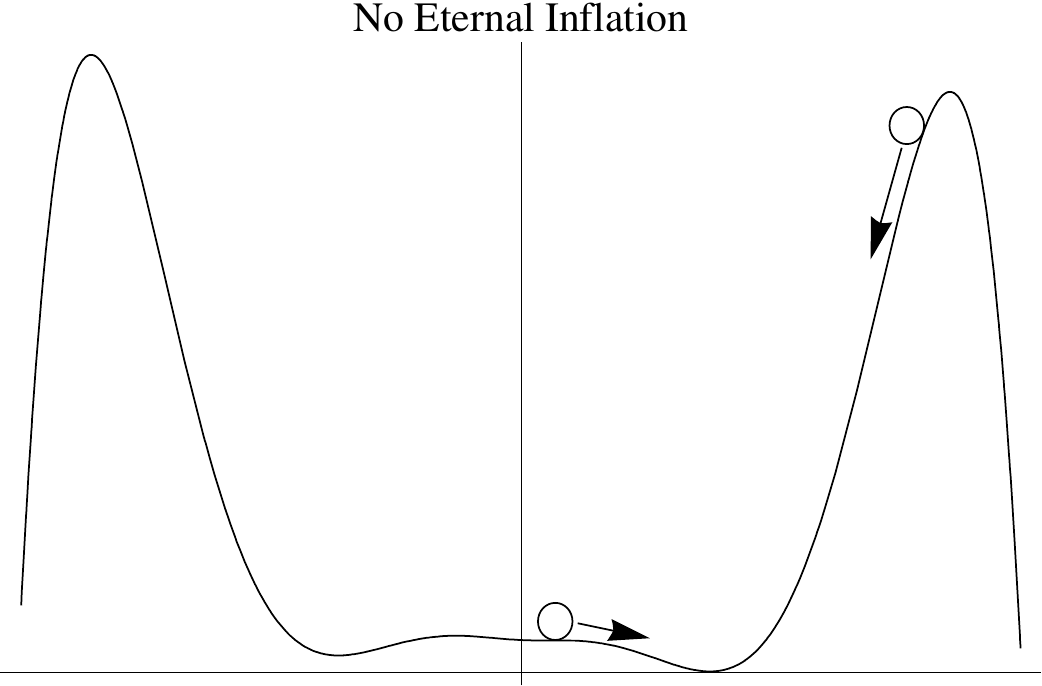}
 \centering
 \caption{We have added a non-renomalizable term to the SC potential Eq.\ \ref{CW}.  
 Hence, the potential has no chaotic eternal inflation.  In addition,
 the potential does not have eternal inflation at the hilltops since the field does not slow roll at the hilltop points.}
 \centering
 \label{noeternfig}
 \efig

SCI also avoids eternal inflation at the hilltop point $\phi_0$.
Near $\phi_0$ we can approximate Eq.~\ref{CWredux} with
\beq
\label{CWsimple}
V(\phi)=\frac{\lambda^\prime}{4}|\phi|^4-\frac{\lambda_{2n}}{2n+4}\frac{|\phi|^{2n+4}}{\Lambda_c^{2n}}
\eeq
where $\lambda^\prime=\lambda \ln(\Lambda_c/\langle\phi\rangle)$ when
 $4n^2 \ln(\Lambda_c/\langle\phi\rangle)>\ln(\lambda/\lambda_{2n})$
 and $\ln(\Lambda_c/\langle\phi\rangle)>1/4$. 
 Then using Eq.~\ref{NoEI},
 SCI will not have eternal inflation at $\phi_0$ if
\beq
\label{NoEIredux}
\frac{|\phi_0|}{\sqrt{2n+4}}<\frac{2}{3} m_{pl}\,\,\,\,\,\,\txt{with}\,\,\,\,\,\,|\phi_0|=
\bigg(\frac{\lambda^\prime}{\lambda_{2n}}\bigg)^{\frac{1}{2n}} \Lambda_c,
\eeq
where $m_{pl}$ is the reduced Planck mass.
As an example, consider a dimension 6 operator and assume the cut off is the reduced Planck mass. 
The non-renormalizable term is then
 equivalent to  Eq.~\ref{gcoupling} i.e.\ a non-minimal coupling to gravity.  
 We then find that there will be no eternal inflation for $\lambda_2>3/8\lambda^\prime\simeq10^{-2}$ or equivalent
in terms of Eq.~\ref{gcoupling} that $\xi< -0.1$.  Also, we have  avoided chaotic inflation since  $|\phi_0|<m_{pl}$.
 Hence, SCI
 can avoid eternal inflation by introducing non-renormalizable terms or by simply coupling the SC field to gravity.  
 
 In fact,  the above scenario is quite general.
$\xi$ is generally not constant and runs with the energy scale of the relevant interactions.
   Hill \& Salopek~\cite{Hill199221} calculated the RGE of $\xi$ for a composite scalar field but they note that their conclusion holds for an
   arbitrary scalar field as well.  They found that $\xi=1/6$  is an IR fixed point.  
   In the UV (or at large field values), $\xi$ can easily  become large and negative, which naturally
   circumvents eternal inflation from happening as shown above.

 In many ways it is not surprising that SCI avoids eternal inflation. Inflation driven by a rolling field requires an unusual situation with 
the necessity of having an incredibly flat potential.  After one has gone to the trouble of having an incredibly flat potential in the first place,
then eternal inflation can occur.  
SCI does not depend upon slow roll dynamics and avoids eternal inflation.

\section{Density Perturbations}\label{perturbations}

The presence of a thermal bath (which is necessary in SuperCool (SC) inflation) will strongly 
suppress curvature perturbations.  
Adiabatic density perturbations go like
\beq
\label{curvpert}
\xi=\frac{\delta\rho}{\rho+p}
\eeq
 where $\delta\rho$ is the variation of density and $\rho$
and $p$ are respectively the average energy density of the Universe and pressure at the time a perturbation
leaves the horizon.  For a discussion on perturbations, see \cite{1990eaun.book.....K} and references there in i.e. \cite{Bardeen:1983qw}.

 As with slow roll inflation, one might imagine introducing a single slowly
 rolling scalar field $\phi_r$ to generate density perturbations. In this case, $\delta\rho\sim$ H V$^\prime(\phi_r)$
   where H is the Hubble parameter when the perturbation leaves the horizon. V($\phi_r$) is the potential of the scalar field. In slow roll, 
   V$^\prime(\phi_r)=-3$H$\dot\phi_r$. The prime refers to partial differential with respect to $\phi_r$, and the
    dot refers to differentiation with respect to time.  $\rho+p$ is summed
  over all matter and energy components.  
 Typically for inflation, $\rho+p=\dot\phi_r^2$, but if there is a thermal
  background then we must include thermal radiation and pressure where $\rho_{r}=1/3p_{r}$.  The thermal
  component will dominate the sum $\rho+p$, which will suppress adiabatic perturbations generated by the rolling
  field. 
  
  One might imagine that thermal variations might  generate perturbations, 
  but the thermal variation across the initial patch is exponentially small.  Equilibrium processes
   wash out any temperature variations across the patch ($\delta \rho_{r}/\rho+p\ll1$), unless there is a process to generate
   perturbations on all scales in the plasma just prior to the start of inflation.
 We will not consider this case any further, but will leave the problem for a separate paper~\cite{doug1}.
 Instead, we will turn to the  aulos mechanism as discussed in the introduction.

  \subsection{ Aulos Mechanism}\label{aulosmechanism}
  
We introduce an explicit  model to implement the  aulos mechanism as outlined in the introduction.
  We introduce a new complex scalar field $\chi_a$ and 2 new scalar fields $\chi_b$ and $\chi_c$.  The new fields have a potential
 \beq
 \label{toymodel}
 \begin{aligned}
 &V_{abc}=-\frac{\mu^2}{2}(\chi_b^2+\chi_c^2)+\frac{\lambda}{4}(\chi_b^2+\chi_c^2)^2\\
 &+\frac{\lambda_\chi}{4}|\chi_a|^4\bigg(\ln\bigg(\frac{|\chi_a|}{\nu}\bigg)-\frac{1}{4}\bigg)+\sum_i^{abc}
  \bigg(\frac{\alpha_i \txt{T}^2}{2} |\chi_i|^2\bigg)\\
  &+\frac{\lambda_b}{2}\bigg(\chi_b^2-\chi_c^2\bigg)|\chi_a|^2-\frac{\lambda_{a\phi}}{2}|\phi|^2 |\chi_a|^2.
  \end{aligned}
 \eeq
 All of the coefficients are positive. 
 There is  an O(2) symmetry between $\chi_b$ and $\chi_c$,
 which is broken by the interaction term with $\chi_a$.   $\chi_a$ has a U(1)$_a$ axial symmetry.
$\chi_b$ and $\chi_c$ get a vev 
when ($\mu>\alpha_b$T$^2$ \& $\mu>\alpha_c$T$^2$).
\footnote{If inflation begins once the temperature of the Universe drops below 100 GeV, then we require
  that $\mu_b$ and $\mu_c$ are $\geq$ 100 GeV
 assuming $\alpha_i$ are order 1 for instance $\alpha=1/4$ for the SC field (See Eq.~\ref{approx}}).

We require that $\chi_a$ is independent of temperature. 
 If the field is temperature dependent then it will be difficult to generate the necessary 
density perturbations for inflation.  We could argue that $\chi_a$ has a zero temperature during inflation. 
Instead, we cancel  the temperature dependent term of
 $\chi_a$ in Eq.~\ref{toymodel} ($\alpha_a$T$^2|\chi_a|^2$) by setting
\beq
\label{cancel}
\alpha_a= \frac{\lambda_b}{\lambda}(\alpha_c-\alpha_b),
\eeq
 which requires a careful selection of the matter coupled to $\chi_a$, $\chi_b$, and $\chi_c$ 
given ($\lambda_b/\lambda$). The coefficients $\alpha_i$ 
in Eq.~\ref{toymodel} depends upon the matter content coupled to $\chi_i$.
We will require that $\chi_b$ and $\chi_c$ have different couplings to fermions i.e.\ $\alpha_c\neq\alpha_b$.
The different couplings to matter will then break the  O(2) symmetry of Eq.~\ref{toymodel} and potentially 
induce a  different $\lambda$ for $\chi_b$ and $\chi_c$ due to quantum corrections.
One can insure that $\chi_b$ and $\chi_c$  do have the same $\lambda$ 
by again carefully choosing the matter content which couples to
$\chi_b$ and$ \chi_c$ such that the loop corrections will be the same for $\chi_b$ and $\chi_c$. 
In addition, we will require that $\lambda_b/\lambda$ is a rational number to 
allow for the cancellation of  ($\alpha_a$T$^2|\chi_a|^2$).  This will require a fair amount of model building.
In a more fundamental theory one might be
able to find such a scenario.
Otherwise naively, Eq.~\ref{cancel} requires severe tuning. 
  We do not address the naturalness of such a scenario at this time.
 
 Regardless, the interaction term, now, cancels the $\alpha_a$T$^2\chi_a^2$ term;
  $\chi_a$ has a Coleman Weinberg potential and then undergoes dimensional transmutation which spontaneously breaks
   U(1)$_a$ i.e. ($\chi_a$ acquires a  vev $\nu$).
With the vev of $\chi_a$ ($\nu\neq0$), we redefine $\chi_a$ in terms of a radial field ($\sigma$) and 
an angular field ($a$) which corresponds to the
pseudo Nambu-Goldstone boson  of $\chi_a=(\nu+\sigma)e^{i a/\nu}$.
 We will associate the pseudo Nambu-Goldstone boson ($a$) with the  aulos field. As one final note, we can avoid
 eternal inflation in the  aulos sector in much the same way that we avoided eternal inflation in the inflaton sector.

\subsection{ Aulos Mass} \label{aulosmass}

\bfig
\includegraphics[width=0.5\textwidth]{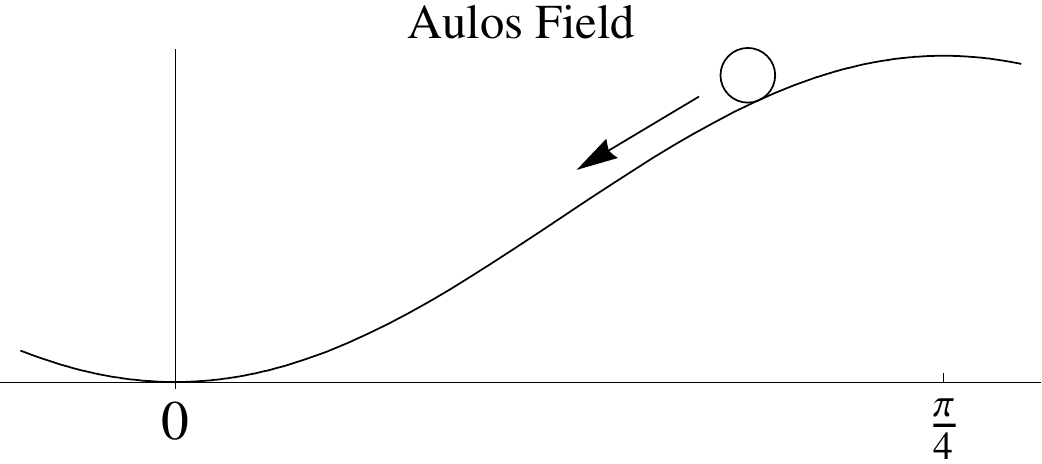}
\caption{Aulos potential Eq.~\ref{aulosPotential} in units of $\nu$. During inflation, the aulos vev and mass are small. At the end of inflation, 
the vev and mass become large.}
\centering
\label{aulosfig}
\efig

We introduce a mass term for the  aulos field ($a$), by softly breaking the U(1)$_a$ symmetry.  The potential of the aulos
field is then,
\beq
\label{aulosPotential}
\begin{aligned}
V(\chi_a)_S&=\frac{\lambda_s}{4}((\chi_a)^4+(\chi_a^\star)^4)\\
&=\frac{\lambda_s (\nu+\sigma)^4}{2}\bigg( \cos\bigg(\frac{4a}{\nu}+\pi\bigg)+1\bigg)
\end{aligned}
\eeq
where $\lambda_s$ is a dimensionless parameter which gives the size of the symmetry breaking 
and $\chi_a^\star$ is the complex conjugate of $\chi_a$. A soft breaking term such as Eq.\ \ref{aulosPotential} can arise from a 
constrained instanton first introduced by t'Hooft~\cite{PhysRevD.14.3432} and more fully developed by Affleck~\cite{Affleck:1980mp}.
We will leave a more careful analysis  for the future and treat Eq.~\ref{aulosPotential} at an effective level.
In Eq.~\ref{aulosPotential}, we have shifted the  aulos field such that
the minimum of the potential is given when $4a/\nu\rightarrow0$ (See Fig.~\ref{aulosfig}).
We might worry about the formation of domain walls. With the given breaking term
 there are 4 degenerate vacua. But the worry is unfounded. 
$\chi_a$ and the  aulos field never reheat at the end of inflation.
 In which case, the Kibble mechanism will not generate domain walls.

 Expanding the cosine at the minimum gives that 
\beq
\label{aulosMass}
m_a^2  =16 \lambda_s \nu^2=128\lambda_s f_a^2 
\eeq
where $m_a$ is the mass of the  aulos field and $f_a=\nu/4$ is the decay constant for the  aulos field. 
The mass is 
proportionate to the spontaneous symmetry breaking scale. 
 Hence, the mass will vary if the spontaneously symmetry breaking scale changes.
\subsection{ Aulos Evolution}\label{aulsoevolution}

 In our toy model during inflation, the  aulos field is Hubble damped with 3H $\simeq10^{-3}$ eV $\geq m_a$.
 The vev of  the SC field is zero and does not contribute to the vev
of $\chi_a$. The vev of $\chi_a$ is initially small.
 At the end of inflation, the SC  field gets a large vev  $\langle\phi\rangle$ which will generate a large and negative
effective mass term for $\chi_a$. Then from Eq.~\ref{toymodel}, $\chi_a$  gets a new large vev ($\nu$) 
\beq
\label{vevafter}
\nu^2\rightarrow \frac{\lambda_{a\phi}\langle\phi\rangle^2}{\lambda_\chi}.
\eeq
The  aulos mass  and decay constant  become large.  See Eq.~\ref{aulosMass}.

When $m_a\geq$3H, the  aulos field begins to oscillate and generates a 	condensate with an energy density 
\beq
\label{aulosdensity}
\rho_a=\frac{1}{3}m_a^2 f_a^2 \theta_0^2=\frac{\lambda_s}{3}\nu^4 \theta_0^2
\eeq
where ($\theta_0=4a/\nu$) is the misalignment  angle of the field when it begins to oscillate. 
 We assume that the field begins to oscillate near the top
of the cosine potential in Eq.\ref{aulosPotential}, in which case ($\theta_0\simeq\pi$).  
We also evaluate $\rho_a$ with the enlarged values of $f_a$ and $m_a$.  We have assumed
that the aulos field barely evolves as $f_a$ and $m_a$ grow to their maximum size. 

At the moment, we have assumed that 
when the  aulos field begins to oscillate that $f_a$ is large and constant, but in fact, $f_a$ is not necessarily initially constant.
  As the barrier trapping the SC field in the false vacuum goes away, it is unclear how the phase transition proceeds. 
  If the transition is first order and the SC field tunnels to the
   true vacuum of the theory, then $f_a$ will in fact be constant.  Otherwise,
  the SC field will experience a period of rolling and potentially oscillating 
  around the minimum of the potential Eq.~\ref{CW}.
  In this later case, if we want to determine precisely the 
  energy density in the  aulos field and SC field, we need to solve
   the coupled differential equations of motion for the  aulos and SC field.
We leave this problem for a future paper. Regardless in our toy model,
 we show that the decay rate of the SC field $\Gamma_\phi$ is
a factor of 10 larger than the decay rate for the  aulos field $\Gamma_a$ 
and can be made much larger with a lighter Z$^\prime$ (See Eq.~\ref{Apexdecay}). 
In addition, in the case of parametric resonance the inflaton will only oscillate once or twice before decaying. 
We can in fact treat $f_a$ as constant during most of the evolution of the  aulos field.

The  aulos field in our model will predominantly decay into photons
 (One could also consider a different decay route but leave that for future work).
We introduce a set of  new chiral fermions charged 
under the global U(1)$_a$ axial symmetry such that we can write down a Yukawa 
coupling between the new chiral fermions and $\chi_a$.  In addition, the new chiral fermions 
are charged under EM, then due to the QED anomaly the  aulos field can decay directly into photons with
\beq
\label{aulosDecay}
\Gamma_{a\rightarrow\gamma\gamma}=\frac{\alpha^{\prime2}}{64 \pi^3}\frac{m^3_a}{f_a^2}
\eeq
 where $\alpha^\prime=g^{\prime2}/4\pi\times N^\prime$. $g^\prime$ is the coupling constant of the  aulos fermions under U(1)$_{\txt{EM}}$
 and $N^\prime$ is a numerical factor which depends upon the number of fermions and their charge under EM.
  We will take 
 $\alpha^\prime\simeq10^{-2} \alpha$ (which is reasonable if the fermions have a fractional charge under EM or couple due to kinetic mixing).
  As long as the fermions are not charged under QCD and are heavy,  
 we can avoid constraints from rare decays and collider physics  with $f_a\geq 100$ GeV and $m_a\geq1$ GeV (See \cite{Essig:2010gu} and
 references within).
 We also require the fermions to be sufficiently heavy to avoid LEP constraints. 
 The fermion's mass should be greater than roughly $100$ GeV.
 When $\chi_a$ acquires a vev, the new chiral fermions become very massive TeV scale for $\order(1)$ Yukawa couplings.

\subsection{ Observational Signatures}\label{signatures}

Quantum fluctuations during inflation cause spatial variation of the misalignment angle $\theta_0$ of the aulos field
 across the Universe. At the end of inflation, the aulos field begins to oscillate, once the mass of the aulos field
 becomes larger than the Hubble parameter.  At which point, variations in the misalignment angle during inflation
induce  an isocurvature perturbation~\cite{Seckel:1985tj}. 

Isocurvature are also
 induced in the QCD axion~\cite{Seckel:1985tj} and the curvaton model but in a different manner from the aulos field.
   The QCD axion and curvaton are
  Hubble damped  during inflation and experience fluctuations in their misalignment angle. Inflation ends; the Universe reheats and then cools.
  Eventually, the Hubble parameter becomes smaller than the mass of the axion or curvaton. 
  The QCD axion and curvaton then begin to oscillate which generate isocurvature perturbations.
   In contrast, the  aulos field begins to oscillate and generate isocurvature perturbations  immediately at the end of inflation
   when the mass of the  aulos field grows and becomes
  larger than the Hubble parameter.

 When the  aulos field decays, the isocurvature perturbation
 becomes a real curvature perturbation~\cite{Mollerach:1989hu}. 
 We can now give an expression for the normalization of the primordial power law spectrum~~\cite{Lyth:2002my}
 \beq
\label{aulospert}
\Delta_{\xi}(k_0)= \frac{r\times q}{3\pi}\bigg( \frac{\txt{H}}{f_a\theta_0}\bigg)\bigg\vert_{\txt{Hor}}
\eeq
where $r=\rho_a/\rho$ is the  ratio of energy density of the  aulos field $\rho_a$  over the total energy density of the
Universe $\rho$ when the  aulos field decays. The ratio of (H$/f_a \theta_0$) is to be evaluated during inflation, where H is the Hubble constant,
$f_a$ is the  aulos decay constant, and $\theta_0$ is the misalignment angle of the aulos field.
  We will take $q=1$,\footnote{$q$ parameterizes nonlinear effect in the 
oscillation of the  aulos field Eq.~\ref{aulosPotential}.  See~\cite{Lyth:2001nq} \cite{Lyth:2002my} for more details.}
for the remainder of the paper.

The power-law index $n_s$ of the primordial power spectrum (following~\cite{Lyth:2002my}) is 
\beq
\label{ns}
n_s\simeq1\pm\frac{2}{3}\frac{m_a^2}{\txt{H}^2}+2\frac{\rho_r}{\Lambda^4}
\eeq
where respectively $(-)$ holds if during inflation the  aulos field is near the top of the potential in Eq.~\ref{aulosPotential} 
and conversely $(+)$ holds if the  aulos field is near the bottom of the potential in Eq.~\ref{aulosPotential} ( 
 WMAP 7 gives that $n_s=0.967$~\cite{Komatsu:2010fb}.  We will assume that the  aulos
field is stuck near the top of the potential and the $(-)$ case holds). In the second term of Eq.~\ref{ns},
$\rho_r$ is the energy density of radiation during inflation over the vacuum energy $\Lambda^4$.  
During inflation the second term in Eq.~\ref{ns} can be safely ignored. 

The  aulos mechanism  can also generate  non-Gaussianities~\cite{Lyth:2002my} with 
\beq
f_{\txt{NL}}=\frac{5}{4r}.
\eeq
WMAP 7 constrains $f_{\txt{NL}}\lesssim100$ which requires that $r\gtrsim10^{-2}$.  
 We will take $r= 3\times10^{-2}$, which gives an $f_{\txt{NL}}=42$. We can reasonably 
 have a much smaller $f_{NL}$ but picked a value which would be observable.
   In the near term, Planck and 21-cm experiment can detect ($f_{\txt{NL}}\simeq5$ to 10)~\cite{PhysRevLett.107.131304}.

In addition, the  aulos field can  isocurvature perturbations. At present, the size of isocurvature perturbations must be less
than about a tenth of the size of adiabatic density  perturbations~\cite{Komatsu:2010fb}.
The isocurvature constraints for the  aulos field are the same as with the curvaton 
(See \cite{Lyth:2002my} for more details). We will just summarize the effects.  Isocurvature constraints require baryogenesis  and 
dark matter (DM) genesis
to occur after the decay of the  aulos. In which case, there will be no isocurvature perturbations.
  
  Instead, if the  aulos field can decay directly into baryons or DM, then the  aulos field could generate interesting levels of
  isocurvature perturbations, which future experiments could detect.
   We will leave the last possibility for future research and will only require that the  aulos field decays before
  the production of DM and baryons.  Below, we show that the  aulos field for the parameters chosen decays almost instantaneously. 
  At that time, 
  the temperature of the 
  Universe is around 500 GeV.  Hence, one can use EW  baryogenesis to generate the baryon asymmetry, 
  which occurs once the temperature of the Universe is around
  100 GeV~\cite{Wainwright:2009mq}.
   As to dark matter, we leave that for a separate paper, but the QCD axion
    would be a viable DM candidate.

\subsection{ Aulos Parameters-Colliders}\label{collider}

We can now relate the parameters of the aulos field to the primordial power spectrum.
The size of the soft breaking term of U(1)$_a$  (See Eq.~\ref{aulosPotential}) 
\beq
\label{lambda_s}
\lambda_s= 6\pi^2 (1-n_s)\bigg(\frac{\Delta_{\xi}(k_0)\,\theta_0}{r \times q}\bigg)^2.
\eeq
can be given in terms of $n_s$, $r$, $q$ and ($\theta_0\simeq\pi$).  $\Delta_{\xi}^2(k_0)=2.43\times10^{-9}$~\cite{Komatsu:2010fb}.  
We find that $\lambda_s=4.3\times10^{-5}$, which is small but technically natural.  
With inflation at the TeV scale and $r=3\times10^{-2}$, Eq.~\ref{aulospert} fixes  $\nu\simeq4\times10^{-2}$ eV 
during inflation. Also, $\lambda_{a\phi}\lesssim1$, such that the  aulos field does not inadvertently destabilize the SC field before
inflation ends.  The small vev $\nu$ generates an effective negative mass term for the SC field. Conversely if $\lambda_{a\phi}\sim1$,
we could then use the small negative mass term to end SCI rather than the Witten mechanism.  
This negative mass case would be similar in spirit to thermal inflation. We will leave this case for future research.

At the end of inflation, $\nu=7$ TeV  and $m_a=730$ GeV with $\lambda_\chi\simeq\lambda_{a\phi}$ 
and $\langle\phi\rangle\simeq 10$ TeV. 
For simplicity, we will  assume that the vacuum energy associated with 
the $\chi_i$ fields is subdominant compared to the scale of inflation ($V_{abc}\lesssim\Lambda^4$), 
which implies that $\lambda_{a\phi}\lesssim10^{-4}$.
In which case, the mass of $\chi_a$ is still heavier than $M_X$. Then, $\phi$ will not decay into $\chi_a$.
One could  imagine a more complicated decay process, but we  will ignore the possibility at the moment.

 Eq.~\ref{aulosDecay} gives the decay rate
 of the  aulos field ($\Gamma_a=2\times10^{-3}$ eV, with $\alpha^\prime=10^{-2} \alpha$), 
 which is long  compared to the decay rate of the SC field ($\Gamma_\phi=0.03$ eV)
and short compared to the Hubble scale at the end of inflation ( H$=4\times10^{-4}$eV).
So in fact, the  aulos and SC field decay almost instantaneously relative to H.

The  aulos field can make connections between collider physics and the CMB.  
We can redefine the mass and the decay rate of the  aulos field in terms of parameters  
determined by cosmological measurements i.e.\ Eq.\ref{lambda_s} and the coupling of  aulos field to fermions. 
SC vev and the  aulos decay constant should  be on the same scale. Due to the low scale of
SCI, colliders could rule out the scenario or potentially give a route of discovery.
 In addition to LHC,
there will be new searches for hidden sectors at Jefferson Lab, JPARC, and future Long Baseline Neutrino experiments
such as LBNE~\cite{Essig:2010xa,Essig:2010gu}. We will leave a detailed discussion of particle physics searches for SCI
 to a future paper~\cite{doug1}.

\section{Conclusion}\label{conclusion}

In conclusion, we have proposed SuperCool (SC) inflation as a solution to old inflation  The Universe
inflates solving the horizon and flatness problems.  At a sufficiently low temperature, a  non-Abelian gauge group becomes strongly
coupled which generates a tadpole term. 
The new term destabilizes the false vacuum, providing a graceful exit.
  Hence, SuperCool Inflation (SCI) generates a successful period of inflation.

 There have been other models proposed which have a graceful exit, 
such as double field inflation~\cite{Adams:1991ma}, a variant of hybrid inflation~\cite{PhysRevD.49.748},
and chain inflation~\cite{Freese:2004vs}~\cite{Freese:2006fk} to name a few. 
These models with the exception of chain inflation use the dynamics of a rolling field to give old inflation a graceful exit.
The rolling models will have eternal inflation and must confront the measure problem unlike SCI.
Furthermore, the rolling models are finely tuned  to generate a sufficient amount of inflation~\cite{Adams:1991ma}.
In SuperCool (SC) inflation, the running of a non-Abelian gauge coupling determines the number of efolds of inflation.   Thus, SCI 
is no more finely tuned than QCD, with the caveat that the ``no bare mass" condition is sensible.

SC inflation is not the only inflationary model, which takes advantage of 
thermal effects such as  warm inflation~\cite{Berera:1995ie} and thermal inflation~\cite{Lyth:1995ka}. 
Warm inflation
modifies the friction term for the equation of motion of the inflaton due to the appearance of a finite temperature during inflation
and should be lumped in with other rolling models.
 Thermal inflation and SC inflation are very similar in spirit.  In thermal inflation, an effective mass 
 from the finite temperature potential as with SCI stabilizes a false vacuum. 
 Eventually a negative mass term destabilizes the false vacuum and the field transitions to the true vacuum. 
Unlike SC inflation, thermal inflation is too brief
 to solve  the flatness and horizon problems and does not have a way to generate density perturbations
  but is sufficiently long to dilute unwanted relics such as gravitinos etc..

As argued in the text, the TeV scale seems to be a natural fit to SC inflation. 
 The scale of SC inflation could be much lower than the TeV scale with a hard lower bound given by BBN ( a few MeV). 
 In principle, there is no upper bound on the scale of SC inflation if one can ignore  non-minimal coupling to
 gravity or non-renormalizable terms in general.  But as argued in the text, one should not neglect for instance a non-minimal coupling
 to gravity.  In which case, there is an upper bound on the scale of SC inflation on the order of 100 TeV.  
 
 TeV scale inflation prompts a question on the naturalness of the initial conditions.  We have assumed as our starting
  point that the initial patch which hosts the visible Universe today was hot, homogeneous, and isotropic when inflation began.
  We do not address the naturalness of such a set up, but the same ambiguity existed in Guth's original model.   
  In that instance, Guth argued that quantum gravity just might generate the right initial conditions.  Even in rolling models
  one still faces the same difficulty or one might hope that eternal inflation offers a route out of initial condition conundrum,
  but then one must address the measure problem.
   At the moment, there is no obvious solution to the initial condition problem in any model of inflation, without invoking
   eternal inflation, which is good or bad depending upon your tastes.

As an interesting feature, SC inflation naturally avoids eternal inflation. As argued in the text, eternal inflation appears both ubiquitous and 
fraught with difficulty.  One solution to eternal inflation is no eternal inflation.  Eternal inflation occurs at hilltop points (the maximum of the potential)  
and at large field values (chaotic inflation). See Fig.\ \ref{eternfig}. Non-minimal coupling to gravity of the
SC field or non-renormalizable
terms in the effective potential for the SC inflaton can prevent chaotic eternal inflation.  At hilltop points,  
the field has no slow roll.  We show in the Appendix that  the field will then not have eternal inflation at the hilltop points.

 In fact, we should not be surprised 
that eternal inflation does not occur for SC inflation.  Eternal inflation for a rolling field is a symptom of slow roll which generally 
requires a large amount of tuning~\cite{Adams:1990pn}. 
Only once we have gone to the trouble to finely tune our potential to get slow roll that we run into eternal
inflation.  SC inflation does not depend upon slow roll. Hence, SC inflation avoids eternal inflation.

One of the great success of slow roll inflation is generating adiabatic density perturbations.  The SC inflaton generates 
the vacuum energy which drives inflation and ends inflation due to the Witten mechanism, but does not generate
appreciable density perturbations.
   We have imagined two possible ways to generate perturbations. First,  if there exist initial perturbations in the plasma just
  before inflation begins, then those perturbations on the largest scales\footnote{
  We only observe perturbations on the largest scales which correspond to perturbations that  leave the horizon 
  within the first 10 efoldings of inflation. On small scales equilibrium processes wash out perturbations, but 
  large scale perturbations can persist.  
  } 
   will redshift as the Universe inflates  and produce 
  the adiabatic density perturbations we see today.  For instance, one might imagine 
  a first order phase transition just before inflation begins. Bubble collisions might
   seed perturbations in the plasma of just the sort we need.
  In this paper we have not addressed this scenario and leave such considerations to a separate paper~\cite{doug1}.

  Instead we introduced a new mechanism to generate 
 density perturbations which we have dubbed the  aulos mechanism.  
 The  aulos field is a pseudo Nambu-Goldstone boson similar to an axion.  
 De Sitter fluctuations induce variations in the misalignment angle of the  aulos field.  At
  the end of inflation the mass of the  aulos field becomes large.  
  At which point, the  aulos field begins to oscillate and generate a condensate.
  The variation in the misalignment angle induces an isocurvature perturbation.  Real adiabatic density perturbations
  are generated when the  aulos field decays into radiation.  As noted previously, the  aulos mechanism successfully 
   matches observations from the CMB and potentially
  can generate non-Gaussianities and isocurvature perturbations, which could be observed by Planck.

  We define an aulos field more broadly as any 
  pseudo Nambu-Goldstone boson which has a small spontaneous symmetry breaking scale and mass  during inflation
  and a large mass and breaking scale at the end of inflation.
   An aulos field has several interesting applications to inflation.
    We only mention a few.
  In a manner similar to  the curvaton scenario, the  aulos mechanism could be used to generate  
  perturbations in an arbitrary inflation scenario.
 In a separate publication~\cite{doug1}, we use the aulos mechanism to generate perturbations in hybrid models at the TeV scale,
 which  avoids previous constraints \cite{Lyth:1999ty}.
  The curvaton scenario requires that the scale of inflation to be above $10^{10}$ GeV. 
  The  aulos mechanism works for an arbitrarily low scale of inflation. 
  Second, the Affleck-Dine mechanism typically requires that the scale of inflation be near $10^{9}$ GeV.
  An  aulos field which carries baryon number could do baryogenesis via an Affleck-Dine 
  mechanism for an arbitrary scale of inflation. 
  One could potentially implement inflation successfully at the MeV scale. 
  At the moment, we know of no models of inflation, which is successful
  with the scale of inflation below 100 GeV!
 
 We can imagine that the  aulos mechanism can  be used beyond inflationary dynamics.
    The  aulos mechanism can be used to generate
  spatial variation of $\alpha$, which has been observed by~\cite{2010arXiv1008.3907W} and  a spatial variation of dark energy.
   In addition, the  aulos mechanism could  generate compensated perturbations,
  which has recently garnered interest. See \cite{2010ApJ...716..907H,2011arXiv1107.1716G} for a discussion. 
   In the last two cases, the  aulos field would not be associated with a field which generates density perturbations during inflation. 
We will consider these possibilities in separate publications.

\bigskip

{\it Acknowledgements}: We would like to thank P. Adshead, B. Bardeen, S. Dodelson,  E. Kolb, J. Frieman, 
R. Feldmann, P. Fox, N. Gnedin, A. Guth, D. Hooper,  J. Kopp, W. Hu, J. Lykken, G. Miller,  E. Neil, A. Stebbins, A. Upadhye,
 E. Weinberg, M. Wyman and A. Zablocki for discussion and feedback.  
 We would like to extend a special thanks to M. Turner and  C. Hill whose criticism profoundly reshaped the final paper. Finally,
 A.  Albrecht's  encouragement, criticism, and insight was a light in the dark.
  This material is based upon work supported in part by the National Science 
  Foundation under Grant No. 1066293 and the hospitality of the Aspen Center for Physics. 
  D. Spolyar is supported by the Department of Energy.

\bibliographystyle{ieeetr}
\bibliography{infRef2}

\begin{thebibliography}{10}

\bibitem{Guth:1980zm}
A.~H. Guth, ``{The Inflationary Universe: A Possible Solution to the Horizon
  and Flatness Problems},'' {\em Phys.Rev.}, vol.~D23, pp.~347--356, 1981.

\bibitem{Linde:1981mu}
A.~D. Linde, ``{A New Inflationary Universe Scenario: A Possible Solution of
  the Horizon, Flatness, Homogeneity, Isotropy and Primordial Monopole
  Problems},'' {\em Phys.Lett.}, vol.~B108, pp.~389--393, 1982.

\bibitem{Albrecht:1982wi}
A.~Albrecht and P.~J. Steinhardt, ``{Cosmology for Grand Unified Theories with
  Radiatively Induced Symmetry Breaking},'' {\em Phys.Rev.Lett.}, vol.~48,
  pp.~1220--1223, 1982.

\bibitem{Adams:1990pn}
F.~C. Adams, K.~Freese, and A.~H. Guth, ``{Constraints on the scalar field
  potential in inflationary models},'' {\em Phys.Rev.}, vol.~D43, pp.~965--976,
  1991.

\bibitem{2011PhRvD..84b3511K}
J.~{Khoury} and G.~E.~J. {Miller}, ``{Towards a cosmological dual to
  inflation},'' {\em \prd}, vol.~84, p.~023511, July 2011.

\bibitem{Khoury:2001wf}
J.~Khoury, B.~A. Ovrut, P.~J. Steinhardt, and N.~Turok, ``{The Ekpyrotic
  universe: Colliding branes and the origin of the hot big bang},'' {\em
  Phys.Rev.}, vol.~D64, p.~123522, 2001.

\bibitem{Lyth:1995ka}
D.~H. Lyth and E.~D. Stewart, ``{Thermal inflation and the moduli problem},''
  {\em Phys.Rev.}, vol.~D53, pp.~1784--1798, 1996.

\bibitem{Guth:1981uk}
A.~H. Guth and E.~J. Weinberg, ``{Cosmological Consequences of a First Order
  Phase Transition in the SU(5) Grand Unified Model},'' {\em Phys.Rev.},
  vol.~D23, p.~876, 1981.

\bibitem{Witten:1980ez}
E.~Witten, ``{Cosmological Consequences of a Light Higgs Boson},'' {\em
  Nucl.Phys.}, vol.~B177, p.~477, 1981.

\bibitem{1997PhRvL..78.1861L}
D.~H. {Lyth}, ``{What Would We Learn by Detecting a Gravitational Wave Signal
  in the Cosmic Microwave Background Anisotropy?},'' {\em Physical Review
  Letters}, vol.~78, pp.~1861--1863, Mar. 1997.

\bibitem{Knox:1992iy}
L.~Knox and M.~S. Turner, ``{Inflation at the electroweak scale},'' {\em
  Phys.Rev.Lett.}, vol.~70, pp.~371--374, 1993.

\bibitem{Lyth:1999ty}
D.~H. Lyth, ``{Constraints on TeV scale hybrid inflation and comments on
  nonhybrid alternatives},'' {\em Phys.Lett.}, vol.~B466, pp.~85--94, 1999.

\bibitem{German:2001tz}
G.~German, G.~G. Ross, and S.~Sarkar, ``{Low-scale inflation},'' {\em Nucl.
  Phys.}, vol.~B608, pp.~423--450, 2001.

\bibitem{Strassler:2006im}
M.~J. Strassler and K.~M. Zurek, ``{Echoes of a hidden valley at hadron
  colliders},'' {\em Phys. Lett.}, vol.~B651, pp.~374--379, 2007.

\bibitem{Bousso:2006ev}
R.~Bousso, ``{Holographic probabilities in eternal inflation},'' {\em Phys.
  Rev. Lett.}, vol.~97, p.~191302, 2006.

\bibitem{Bousso:2010yn}
R.~Bousso, B.~Freivogel, S.~Leichenauer, and V.~Rosenhaus, ``{Eternal inflation
  predicts that time will end},'' {\em Phys.Rev.}, vol.~D83, p.~023525, 2011.

\bibitem{Nomura:2011dt}
Y.~Nomura, ``{Physical Theories, Eternal Inflation, and Quantum Universe},''
  2011.

\bibitem{Harlow:2011az}
D.~Harlow, S.~Shenker, D.~Stanford, and L.~Susskind, ``{Eternal Symmetree},''
  2011.
\newblock * Temporary entry *.

\bibitem{Garriga:2008ks}
J.~Garriga and A.~Vilenkin, ``{Holographic Multiverse},'' {\em JCAP},
  vol.~0901, p.~021, 2009.

\bibitem{Bousso:2010im}
R.~Bousso, B.~Freivogel, S.~Leichenauer, and V.~Rosenhaus, ``{Geometric origin
  of coincidences and hierarchies in the landscape},'' {\em Phys. Rev.},
  vol.~D84, p.~083517, 2011.

\bibitem{Linde:2008xf}
A.~D. Linde, V.~Vanchurin, and S.~Winitzki, ``{Stationary Measure in the
  Multiverse},'' {\em JCAP}, vol.~0901, p.~031, 2009.

\bibitem{DeSimone:2008if}
A.~De~Simone {\em et~al.}, ``{Boltzmann brains and the scale-factor cutoff
  measure of the multiverse},'' {\em Phys. Rev.}, vol.~D82, p.~063520, 2010.

\bibitem{Vilenkin:2011yx}
A.~Vilenkin, ``{Holographic multiverse and the measure problem},'' {\em JCAP},
  vol.~1106, p.~032, 2011.
\newblock * Temporary entry *.

\bibitem{Mollerach:1989hu}
S.~Mollerach, ``{ISOCURVATURE BARYON PERTURBATIONS AND INFLATION},'' {\em
  Phys.Rev.}, vol.~D42, pp.~313--325, 1990.

\bibitem{Lyth:2001nq}
D.~H. Lyth and D.~Wands, ``{Generating the curvature perturbation without an
  inflaton},'' {\em Phys.Lett.}, vol.~B524, pp.~5--14, 2002.

\bibitem{Lyth:2002my}
D.~H. Lyth, C.~Ungarelli, and D.~Wands, ``{The Primordial density perturbation
  in the curvaton scenario},'' {\em Phys.Rev.}, vol.~D67, p.~023503, 2003.

\bibitem{doug1}
work~in progress

\bibitem{Gildener:1976ih}
E.~Gildener and S.~Weinberg, ``{Symmetry Breaking and Scalar Bosons},'' {\em
  Phys.Rev.}, vol.~D13, p.~3333, 1976.

\bibitem{Coleman:1973jx}
S.~R. Coleman and E.~J. Weinberg, ``{Radiative Corrections as the Origin of
  Spontaneous Symmetry Breaking},'' {\em Phys.Rev.}, vol.~D7, pp.~1888--1910,
  1973.

\bibitem{Dolan:1973qd}
L.~Dolan and R.~Jackiw, ``{Symmetry Behavior at Finite Temperature},'' {\em
  Phys.Rev.}, vol.~D9, pp.~3320--3341, 1974.

\bibitem{Weinberg:1974hy}
S.~Weinberg, ``{Gauge and Global Symmetries at High Temperature},'' {\em
  Phys.Rev.}, vol.~D9, pp.~3357--3378, 1974.

\bibitem{Guth:1980zk}
A.~H. Guth and E.~J. Weinberg, ``{A COSMOLOGICAL LOWER BOUND ON THE HIGGS BOSON
  MASS},'' {\em Phys.Rev.Lett.}, vol.~45, p.~1131, 1980.

\bibitem{Linde:1981zj}
A.~D. Linde, ``{Decay of the False Vacuum at Finite Temperature},'' {\em
  Nucl.Phys.}, vol.~B216, p.~421, 1983.

\bibitem{Hawking:1981fz}
S.~Hawking and I.~Moss, ``{Supercooled Phase Transitions in the Very Early
  Universe},'' {\em Phys.Lett.}, vol.~B110, p.~35, 1982.

\bibitem{Guth:1982pn}
A.~H. Guth and E.~J. Weinberg, ``{Could the Universe Have Recovered from a Slow
  First Order Phase Transition?},'' {\em Nucl.Phys.}, vol.~B212, p.~321, 1983.

\bibitem{PhysRevD.24.1699}
M.~Sher, ``Importance of the coupling-constant temperature dependence in
  supercooled phase transitions,'' {\em Phys. Rev. D}, vol.~24, pp.~1699--1701,
  Sep 1981.

\bibitem{Turner:1992tz}
M.~S. Turner, E.~J. Weinberg, and L.~M. Widrow, ``{Bubble nucleation in first
  order inflation and other cosmological phase transitions},'' {\em Phys.Rev.},
  vol.~D46, pp.~2384--2403, 1992.

\bibitem{1997MPLA...12.2287L}
D.~F. {Litim}, C.~{Wetterich}, and N.~{Tetradis}, ``{Nonperturbative Analysis
  of the Coleman-Weinberg Phase Transition},'' {\em Modern Physics Letters A},
  vol.~12, pp.~2287--2308, 1997.

\bibitem{2009arXiv0909.4541W}
J.~D. {Wells}, ``{Lectures on Higgs Boson Physics in the Standard Model and
  Beyond},'' {\em ArXiv e-prints}, Sept. 2009.

\bibitem{Holdom:1985ag}
B.~Holdom, ``{Two U(1)'s and Epsilon Charge Shifts},'' {\em Phys.Lett.},
  vol.~B166, p.~196, 1986.

\bibitem{Lee:1977yc}
B.~W. Lee, C.~Quigg, and H.~Thacker, ``{The Strength of Weak Interactions at
  Very High-Energies and the Higgs Boson Mass},'' {\em Phys.Rev.Lett.},
  vol.~38, pp.~883--885, 1977.

\bibitem{Lee:1977eg}
B.~W. Lee, C.~Quigg, and H.~Thacker, ``{Weak Interactions at Very
  High-Energies: The Role of the Higgs Boson Mass},'' {\em Phys.Rev.},
  vol.~D16, p.~1519, 1977.

\bibitem{Gunion:1989we}
J.~F. Gunion, H.~E. Haber, G.~L. Kane, and S.~Dawson, ``{THE HIGGS HUNTER'S
  GUIDE},'' {\em Front.Phys.}, vol.~80, pp.~1--448, 2000.

\bibitem{Wainwright:2009mq}
C.~Wainwright and S.~Profumo, ``{The Impact of a strongly first-order phase
  transition on the abundance of thermal relics},'' {\em Phys.Rev.}, vol.~D80,
  p.~103517, 2009.

\bibitem{2011PhRvD..83g5012A}
E.~{Accomando}, A.~{Belyaev}, L.~{Fedeli}, S.~F. {King}, and
  C.~{Shepherd-Themistocleous}, ``{Z$^{'}$ physics with early LHC data},'' {\em
  \prd}, vol.~83, pp.~075012--+, Apr. 2011.

\bibitem{Appelquist:2002mw}
T.~Appelquist, B.~A. Dobrescu, and A.~R. Hopper, ``{Nonexotic neutral gauge
  bosons},'' {\em Phys.Rev.}, vol.~D68, p.~035012, 2003.

\bibitem{Affleck:1984fy}
I.~Affleck and M.~Dine, ``{A New Mechanism for Baryogenesis},'' {\em
  Nucl.Phys.}, vol.~B249, p.~361, 1985.

\bibitem{Dine:2003ax}
M.~Dine and A.~Kusenko, ``{The Origin of the matter - antimatter asymmetry},''
  {\em Rev.Mod.Phys.}, vol.~76, p.~1, 2004.

\bibitem{Banks:1996ea}
T.~Banks and M.~Dine, ``{The Cosmology of string theoretic axions},'' {\em
  Nucl.Phys.}, vol.~B505, pp.~445--460, 1997.

\bibitem{Callan:1970ze}
J.~Callan, Curtis~G., S.~R. Coleman, and R.~Jackiw, ``{A New improved energy -
  momentum tensor},'' {\em Annals Phys.}, vol.~59, pp.~42--73, 1970.

\bibitem{Abbott:1981rg}
L.~Abbott, ``{GRAVITATIONAL EFFECTS ON THE SU(5) BREAKING PHASE TRANSITION FOR
  A COLEMAN-WEINBERG POTENTIAL},'' {\em Nucl.Phys.}, vol.~B185, p.~233, 1981.

\bibitem{Hill199221}
C.~T. Hill and D.~S. Salopek, ``Calculable nonminimal coupling of composite
  scalar bosons to gravity,'' {\em Annals of Physics}, vol.~213, no.~1, pp.~21
  -- 30, 1992.

\bibitem{1990eaun.book.....K}
E.~W. {Kolb} and M.~S. {Turner}, {\em {The early universe.}}
\newblock 1990.

\bibitem{1986PhLB..175..395L}
A.~D. {Linde}, ``{Eternally existing self-reproducing chaotic inflanationary
  universe},'' {\em Physics Letters B}, vol.~175, pp.~395--400, Aug. 1986.

\bibitem{Baumann:2007np}
D.~Baumann, A.~Dymarsky, I.~R. Klebanov, L.~McAllister, and P.~J. Steinhardt,
  ``{A Delicate universe},'' {\em Phys.Rev.Lett.}, vol.~99, p.~141601, 2007.

\bibitem{Guth:2000ka}
A.~H. Guth, ``{Inflation and eternal inflation},'' {\em Phys.Rept.}, vol.~333,
  pp.~555--574, 2000.
\newblock David Schramm Memorial Volume.

\bibitem{Freivogel:2011eg}
B.~Freivogel, ``{Making predictions in the multiverse},'' {\em
  Class.Quant.Grav.}, vol.~28, p.~204007, 2011.
\newblock * Temporary entry *.

\bibitem{Bardeen:1983qw}
J.~M. Bardeen, P.~J. Steinhardt, and M.~S. Turner, ``{Spontaneous Creation of
  Almost Scale - Free Density Perturbations in an Inflationary Universe},''
  {\em Phys.Rev.}, vol.~D28, p.~679, 1983.

\bibitem{PhysRevD.14.3432}
G.~'t~Hooft, ``Computation of the quantum effects due to a four-dimensional
  pseudoparticle,'' {\em Phys. Rev. D}, vol.~14, pp.~3432--3450, Dec 1976.

\bibitem{Affleck:1980mp}
I.~Affleck, ``{On Constrained Instantons},'' {\em Nucl.Phys.}, vol.~B191,
  p.~429, 1981.

\bibitem{Essig:2010gu}
R.~Essig, R.~Harnik, J.~Kaplan, and N.~Toro, ``{Discovering New Light States at
  Neutrino Experiments},'' {\em Phys.Rev.}, vol.~D82, p.~113008, 2010.

\bibitem{Seckel:1985tj}
D.~Seckel and M.~S. Turner, ``{Isothermal Density Perturbations in an Axion
  Dominated Inflationary Universe},'' {\em Phys.Rev.}, vol.~D32, p.~3178, 1985.

\bibitem{Komatsu:2010fb}
E.~Komatsu {\em et~al.}, ``{Seven-Year Wilkinson Microwave Anisotropy Probe
  (WMAP) Observations: Cosmological Interpretation},'' {\em
  Astrophys.J.Suppl.}, vol.~192, p.~18, 2011.

\bibitem{PhysRevLett.107.131304}
S.~Joudaki, O.~Dor\'e, L.~Ferramacho, M.~Kaplinghat, and M.~G. Santos,
  ``Primordial non-gaussianity from the 21 cm power spectrum during the epoch
  of reionization,'' {\em Phys. Rev. Lett.}, vol.~107, p.~131304, Sep 2011.

\bibitem{Essig:2010xa}
R.~Essig, P.~Schuster, N.~Toro, and B.~Wojtsekhowski, ``{An Electron Fixed
  Target Experiment to Search for a New Vector Boson A' Decaying to e+e-},''
  {\em JHEP}, vol.~02, p.~009, 2011.

\bibitem{Adams:1991ma}
F.~C. Adams and K.~Freese, ``{Double field inflation},'' {\em Phys.Rev.},
  vol.~D43, pp.~353--361, 1991.

\bibitem{PhysRevD.49.748}
A.~Linde, ``Hybrid inflation,'' {\em Phys. Rev. D}, vol.~49, pp.~748--754, Jan
  1994.

\bibitem{Freese:2004vs}
K.~Freese and D.~Spolyar, ``{Chain inflation: 'Bubble bubble toil and
  trouble'},'' {\em JCAP}, vol.~0507, p.~007, 2005.

\bibitem{Freese:2006fk}
K.~Freese, J.~T. Liu, and D.~Spolyar, ``{Chain inflation via rapid tunneling in
  the landscape},'' 2006.

\bibitem{Berera:1995ie}
A.~Berera, ``{Warm inflation},'' {\em Phys.Rev.Lett.}, vol.~75, pp.~3218--3221,
  1995.

\bibitem{2010arXiv1008.3907W}
J.~K. {Webb}, J.~A. {King}, M.~T. {Murphy}, V.~V. {Flambaum}, R.~F. {Carswell},
  and M.~B. {Bainbridge}, ``{Evidence for spatial variation of the fine
  structure constant},'' {\em ArXiv e-prints}, Aug. 2010.

\bibitem{2010ApJ...716..907H}
G.~P. {Holder}, K.~M. {Nollett}, and A.~{van Engelen}, ``{On Possible Variation
  in the Cosmological Baryon Fraction},'' {\em \apj}, vol.~716, pp.~907--913,
  June 2010.

\bibitem{2011arXiv1107.1716G}
D.~{Grin}, O.~{Dor{\'e}}, and M.~{Kamionkowski}, ``{Do baryons trace dark
  matter in the early universe?},'' {\em ArXiv e-prints}, July 2011.

\end{thebibliography}

\section{Appendix}

We show that if slow roll conditions are violated, then there can be no eternal inflation near hilltop (local maximum)
points $\phi_0$  with  
\beq
V^\prime(\phi_0)=0 \txt{ and } V^{\prime\prime}(\phi_0)<0.
\eeq  
where $\partial_\phi V(\phi_0)=V^\prime(\phi_0)$.

Suppose there exists a small patch of space inflating.
The patch  has a field value near  $\phi_0$.  We can approximate the potential $V(\phi)$ near $\phi_0$
with a Taylor series
\beq
\label{taylor}
V(\phi)\simeq \Lambda^4-\frac{M^2}{2} (\phi-\phi_0)^2
\eeq
where $\Lambda^4=V(\phi_0)$ and $M^2=|V^{\prime\prime}(\phi)|$. 
We can now simplify our analysis by just shifting the hilltop point to the origin. Set $\phi_0=0$, then
\beq
\label{simp}
V(\phi)\simeq \Lambda^4-\frac{M^2}{2} (\phi)^2.
\eeq
We can solve the equation of motion $\ddot\phi+3\txt{H}\dot\phi-M^2\phi=0$ when slow roll is violated
with $M>3\txt{H}$ by neglecting the Hubble term.
We then find that
\beq
\label{eqm}
\phi(t)\simeq\phi_0e^{Mt}.
\eeq
We can see that for $\phi_0=0$, the field is stuck at the origin. The field would not evolve and the Universe would eternally inflate,
but even an infinitesimal quantum fluctuation would kick the field to a nonzero field value where it could roll away and stop inflating. 

A proxy for the fraction of the Universe which is eternally inflating at the origin at some time $t$ is then just  
 \beq
 \label{originInflation}
 P_e=e^{3\txt{H}t} p(t)_{non}
 \eeq
 where H is the Hubble parameter H$=\Lambda^2/m_{pl}$, and $p(t)_{non}$ is the probability of not having a quantum fluctuation.
 If $p(t)_{non}$ grows slower than the volume factor $e^{3\txt{H}t}$, the fraction of the Universe inflating at the origin becomes exponentially large. Conversely if the probability function grows faster than the volume factor, then  
 there is no eternal inflation.
 The De Sitter fluctuations are Gaussian in Eq.\ref{originInflation},
  in which case, $p(t)_{non}=(1-$Erf$(t))\sim e^{-t^2}/t$. Hence, the fraction of the Universe which
 inflates at the origin exponentially goes to zero with increasing time.

We now look for a region in which the Universe  can eternally inflates. 
   Presumably if Eq.~\ref{simp}
  admits eternal inflation, then there must exist a region $|\epsilon|>0$ where quantum fluctuations keep the Universe inflating.
  The region cannot extend out to infinity. 
   The vacuum energy goes to zero and then negative for sufficiently large field values.
   Inflation only will occur if the vacuum energy is positive.
    Hence, we will require
  that the vacuum energy density stay positive.  This requires that  we will only consider fluctuations  to points which are
  at most $\sqrt{2}m_{pl}/3$ away from the origin.  
   Hence,  if eternal inflation occurs, there must exist a region $0<|\epsilon|<\sqrt{2}m_{pl}/3$ where
  quantum fluctuations keep the field within $\epsilon$ of the origin.  Hence, we will only consider fluctuations which take the field to values 
  within $\epsilon$ of the origin.  If we had chosen badly and are interested in fluctuations to values larger than $\epsilon$, we can just redefine
  $\epsilon$ to include the larger values.  
  Hence by searching for $\epsilon$, we can determine if the potential eternally inflates. 

 We consider a stochastic process of rolling and quantum kicks to characterize the evolution of the field during eternal inflation.  
 First, assume that a point $\epsilon$ exists. Then, there exists a region within $\epsilon$ of the origin in which eternal inflation can occur. 
 Set the field value initially to $\epsilon$, where we remind the reader that
  $0<|\epsilon|<m_{pl}/3$.
 At time $t$, the field has rolled to $\Delta\phi=\phi(t)$ (See Eq.\ref{eqm}), and
 a quantum fluctuation kicks the field to a point $\delta$ such that $|\delta|\leq|\epsilon|$. 
  The field then begins to roll to a new point $\Delta\phi^\prime$ where quantum fluctuations kick the field to a new point $\delta^\prime$
  such that $|\delta^\prime|\leq\epsilon$. The process continues to repeat itself.  And a fraction of the Universe continues to inflate.

 We consider the first such cycle of roll and quantum kick as discussed above.
After one cycle, our proxy for the fraction of the Universe inflating with a field value within $\epsilon$ of the origin is
 \beq
 \label{prox2}
 P_e(1)=e^{3Ht}p(t)_t p(\Delta\phi,\delta)_\Delta
 \eeq
 where
 \beq
 \label{pt}
 p(t)_t=\frac{1}{\sqrt{2\pi}}\frac{\txt{H}}{2\pi}\int^t_0 e^{{\textstyle-\frac{1}{2}(\frac{t^\prime\txt{H}}{2\pi})^2}}dt^\prime
\eeq
  is the probability of having a quantum fluctuation at a time $t$
  and 
  \beq
\label{pd}
p(\Delta\phi,\delta)_\Delta=\sqrt{ \frac{2\pi}{\txt{H}^2}}\int^{\Delta\phi+\delta}_{\Delta\phi-|\delta|}d\phi e^{-\frac{1}{2}{\textstyle( \frac{\phi}{\txt{H}})^2}}
\eeq 
is the probability of 
 having a quantum fluctuation from the point $\Delta\phi$ to within $\delta$ of the origin.

By considering different cases, we can analytically  show that $P_e(1)<1$, for any $\epsilon$.
In the first case, take $t>(H/2\pi)^{-1}$, then $p(t)_t\lesssim1$ and $\Delta\phi\gg\epsilon>|\delta|$. 
 We can approximate Eq.~\ref{pd} with
 \beq
 p(\Delta\phi,\delta)_\Delta\lesssim\frac{2|\delta|}{\sqrt{2\pi \txt{H}^2}} e^{-\frac{1}{2}{\textstyle( \frac{2\pi(\Delta\phi-|\delta|)}{\txt{H}})^2}}
 \eeq
 We note that $|\delta|=\epsilon e^{-M\Delta t}=\Delta\phi e^{-M(t+\Delta t)}$ where $\Delta t$ is the time it would take the field to roll from $\delta$
 back to $\epsilon$. 
 After making the appropriate substitutions and maximizing the probability,
 \beq
 P_e(1)\lesssim\sqrt{\frac{2}{\pi}} e^{{\textstyle(3H-M)t-(1-\upsilon)^2}} e^{\textstyle-M\Delta t}<1
 \eeq 
 where $\upsilon\equiv\delta/\Delta\phi<1$.

Second, consider (3H)$^{-1}<t\leq 2\pi/$H, then 
\beq
\label{pts}
p(t)\lesssim\frac{1}{\sqrt{2\pi}}\frac{\txt{H}}{2\pi}t <1
\eeq
and
\beq
\label{>3H}
 P_e(1)\lesssim\frac{1}{\pi} e^{{\textstyle(3H-M)t-(1-\upsilon)^2}} e^{\textstyle-M\Delta t} \frac{\txt{H} t}{2\pi}<1
 \eeq
Third let $t<(3\txt{H})^{-1}$,  and  $\Delta\phi-\delta>$H$/2\pi$.  The result is the same as before Eq.~\ref{>3H}.

Finally, consider $t<(3\txt{H})^{-1}$ and  $\Delta\phi-\delta<$H$/2\pi$.  In this case, we are looking for fluctuations which take the field
to points between $\delta$ and $\epsilon$ and not within $\delta$ of the origin.  Thus, the limits of Eq.~\ref{pd} are different in this case. 
We find that 
\beq
\label{pd2}
p(\Delta\phi,\delta)\lesssim\sqrt{2\pi}\, \frac{(\epsilon-\delta)}{\txt{H}}<1
\eeq  
and
 \beq
 \label{<3H}
P_e(1)\lesssim \frac{e^{3H t}}{{2\pi}} t \epsilon (1-e^{-M\Delta t})<1
\eeq

Hence after one cycle, the fraction of the Universe eternally inflating has decreased regardless of $\epsilon$.
Hence by contradiction, a point $\epsilon$
does not exist. Thus, there is no eternal inflation.  In fact as we iterate the stochastic process of rolling an quantum kicks, 
 the fraction of the Universe eternally inflating becomes exponentially suppressed. Thus in the case that,
slow roll is violated ($M>3$H), we do not find eternal inflation near the hilltop of our potential.

 Conversely if we consider the slow roll case ($M<3$H),
then the solution to the equation of motion has the same form as Eq.~\ref{eqm} but we simply send $M\rightarrow M^2/3H$.  Upon
analyzing Eq.~\ref{prox2} in the slow roll case, one clearly finds eternal inflation.

\end{document}